\documentclass[3p,twocolumn]{elsarticle}
\usepackage[utf8]{inputenc}
\usepackage{amsmath,amssymb}
\usepackage{hyperref}
\usepackage{caption}

\usepackage{longtable}
\usepackage{threeparttable}
\usepackage{xcolor}
\usepackage[linewidth=1pt]{mdframed}
\newcommand{\printtext}{true}
\newcommand{\printnotes}{false}
\newcommand{\printtables}{true}
\newcommand{\printfigures}{true}
\newcommand{\printauthors}{true}

\usepackage{ifthen}

\begin{document}

  \title{Emissions Reporting Maturity Model: supporting cities to leverage emissions-related processes through performance indicators and artificial intelligence\tnoteref{t1}}
  \tnotetext[t1]{This document presents the results of a master's dissertation in emissions reporting theme.}

  \ifthenelse{\equal{\printauthors}{true}}
  {
  
  \author[1]{Victor de A. Xavier\corref{cor1}}
  \ead{victorx@cos.ufrj.br}
  \cortext[cor1]{Corresponding author}
  \author[1,2]{Felipe M.G. França}
  \ead{felipe@ieee.br}
  \author[1]{Priscila M.V. Lima}
  \ead{priscilamvl@cos.ufrj.br}
  
  \affiliation[1]{organization={Universidade Federal do Rio de Janeiro},
    addressline={Av. Pedro Calmon, 550 - Cidade Universitaria},
    postcode={21941-901},
    city={RJ},
    country={Brazil}}

  \affiliation[2]{organization={Instituto de Telecomunicacoes - Universidade do Porto},
    addressline={Rua Dr. Roberto Frias, s/n},
    postcode={4200-465},
    city={Porto},
    country={Portugal}}

  }

  \date{December 2023}

  \frontmatter

  \begin{abstract}
  Climate change and global warming have been trending topics worldwide since the Eco-92 conference. However, little progress has been made in reducing greenhouse gases (GHGs). The problems and challenges related to emissions are complex and require a concerted and comprehensive effort to address them. Emissions reporting is a critical component of GHG reduction policy and is therefore the focus of this work.

  It is crucial to improve the process efficiency of emissions reporting in order to achieve better emissions reduction results, as there is a direct link between effective emissions policies implemented by cities and emissions reduction (or increase) due to the effectiveness of these policies. Hence, to achieve this goal, this work proposes a series of steps to investigate, search and develop performance indicators (PIs) for emissions reporting. These performance indicators are based on the data provided by cities on the processes they go through to address emission problems. PIs can be used to guide and optimize the policies responsible for implementing emission reduction measures at the city level. Therefore, the main goal of this work is two-fold: (i) to propose an emission reporting evaluation model to leverage emissions reporting overall quality and (ii) to use artificial intelligence (AI) to support the initiatives that improve emissions reporting.
  
  Thus, this work presents an Emissions Reporting Maturity Model (ERMM) for examining, clustering, and analysing data from emissions reporting initiatives to help the cities to deal with climate change and global warming challenges. The model is built using Capability Maturity Model (CMM) concepts and uses artificial intelligence clustering technologies, performance indicator candidates and a qualitative analysis approach to find the data flow along the emissions-related processes implemented by cities. The Performance Indicator Development Process (PIDP) proposed in this work provides ways to leverage the quality of the available data necessary for the execution of the evaluations identified by the ERMM. Hence, the PIDP supports the preparation of the data from emissions-related databases, the classification of the data according to similarities highlighted by different clustering techniques, and the identification of performance indicator candidates, which are strengthened by a qualitative analysis of selected data samples. 
  
  Thus, the main goal of ERRM is to evaluate and classify the cities regarding the emission reporting processes, pointing out the drawbacks and challenges faced by other cities from different contexts, and at the end to help them to leverage the underlying emissions-related processes and emissions mitigation initiatives. 
  
  \end{abstract}


  \maketitle

  \section{Introduction}

  \ifthenelse{\equal{\printnotes}{true}}
  {
  \begin{mdframed}[leftmargin=1pt,rightmargin=1pt]
  \color{red}
  \begin{itemize}
  \item Climate Change and Global Warming..
  \subitem Climate Change and Global Warming definition..
  \subitem IPCC alerts..
  \subitem Other institutes alerts..
  \item Greenhouse Effect Gases (GHG) Emissions
  \subitem Emissions contributions to CCGM..
  \subitem Emissions reduction state of the art..
  \subitem Emissions indicators..
  \subitem Emissions monitoring initiatives..
  \item GHG Impacts Mitigation Initiatives..
  \subitem CDP Initiative..
  \subitem GCOM Initiative and others..
  \subitem ? Cities are capable of emissions reduction policymaking based on available data ? 
  \item Goals..
  \subitem Emissions Reporting Analysis using AI..
  \subsubitem Overview of Emissions Reporting Analysis process..
  \subsubitem Overview of AI techniques used in the process..
  \subsubitem Overview of Maturity Models concepts.. 
  \subitem To help cities to leverage emissions reporting processes..
  \item Contributions..
  \subitem Performance Indicators Development Process..
  \subitem Emissions Reporting Maturity Model..
  \subitem Emissions Reporting Maturity Level..
  \subsubitem ERMM in action with 814 cities from CDP database..
  \end{itemize}
  \end{mdframed}
  }
 
  \ifthenelse{\equal{\printtext}{true}}
  {

  The United Nations Climate Change Conference (COP26)\footnote{https://ukcop26.org/} hosted by the United Kingdom in Glasgow has finished last thirteenth of November with a clear message: time is running out and world leaders must commit to actions than to promises. According to the Intergovernmental Panel on Climate Change (IPCC)\footnote{https://www.ipcc.ch}, the United Nations body responsible for the scientific assessment of climate change, which has produced a special report on the impacts of global warming and associated global greenhouse gas emission pathways\footnote{https://www.ipcc.ch/sr15/}, it is imperative to implement the necessary actions to keep the increase in global average temperature to 1.5 degrees Celsius above pre-industrial levels. The report was prepared in response to the Paris Agreement \footnote{https://unfccc.int} proposals and highlights the implications by comparing the two scenarios of 1.5 and 2 degrees Celsius, as well as the mitigation alternatives that can be applied as part of a global effort.

  The COP26 summit brought parties together to accelerate action towards the goals of the Paris Agreement and the UN Framework Convention on Climate Change. Leading organisations involved in climate change research, policymaking and education such as the International Science Council (ISC)\footnote{https://council.science/}, U.S. Environmental Protection Agency (EPA)\footnote{https://www.epa.gov/}, World Climate Research Programme (WCRP)\footnote{https://www.wcrp-climate.org/}, all point in the same direction. 

  }
  
  \subsection{Greenhouse Gases (GHG) emissions}

  \ifthenelse{\equal{\printnotes}{true}}
  {
  \begin{mdframed}[leftmargin=10pt,rightmargin=10pt]
  \color{red}
  \begin{itemize}
  \item Emissions contributions to CCGM..
  \item Emissions reduction state of the art..
  \item Emissions indicators..
  \item Emissions monitoring initiatives..
  \end{itemize}
  \end{mdframed}
  }

  \ifthenelse{\equal{\printtext}{true}}
  {
  
  Greenhouse gas emissions contribute significantly to the rise in global temperature. For this reason, reducing emissions of these gases should be a central component of strategies to mitigate global warming and the effects of climate change. Figure \ref{fig:figure_co2emissions_map_2019} illustrates how greenhouse gas emissions are distributed globally by looking at the emission totals of the main gas (CO\textsubscript{2}).

  }
  
  \ifthenelse{\equal{\printfigures}{true}}
  {
  \begin{figure}[ht]
  \centering
  \includegraphics[width=\columnwidth]{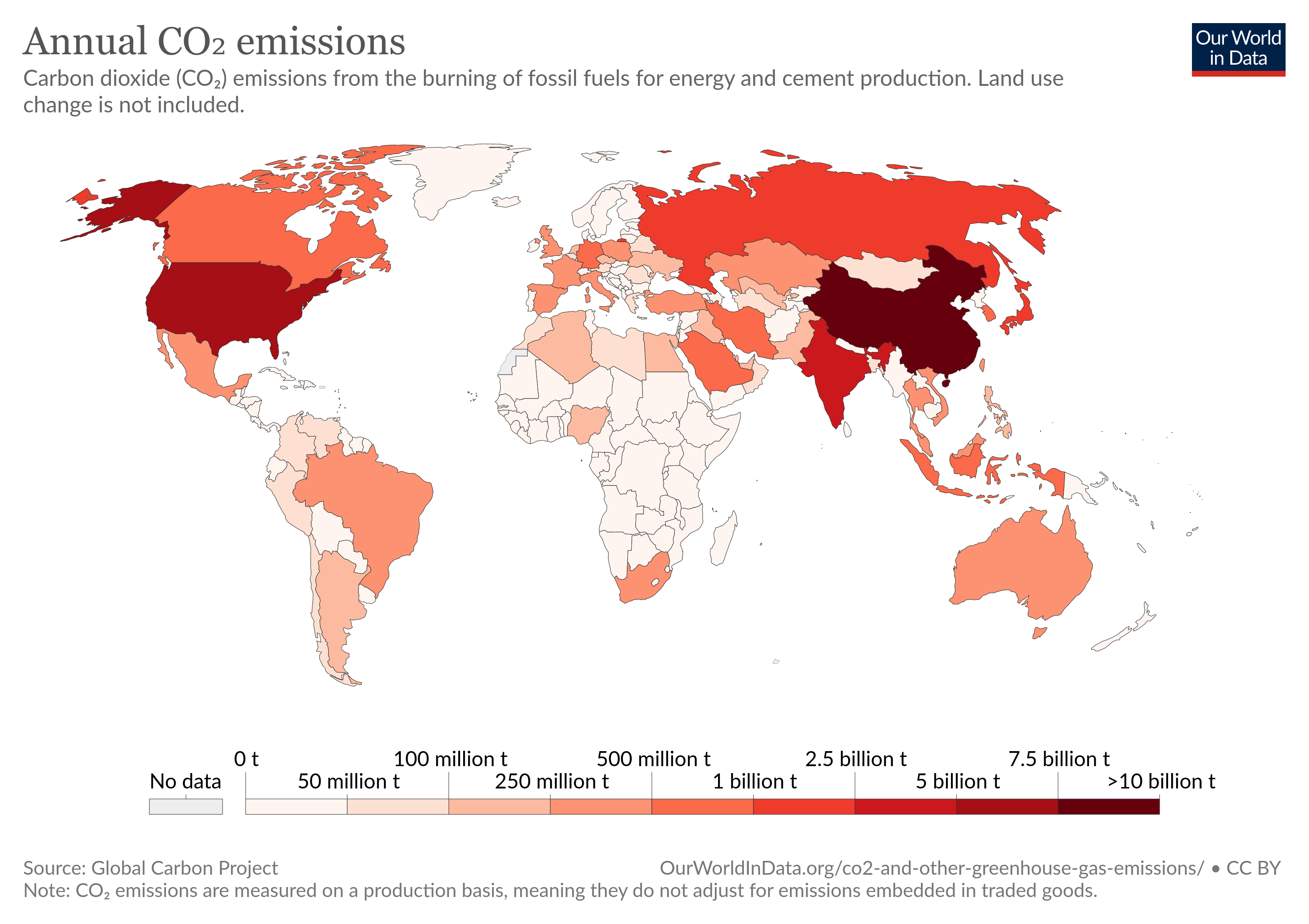}
  \caption[CO\textsubscript{2} emissions world map 2019]{CO\textsubscript{2} emissions world map 2019. Source: \cite{owidco2andgreenhousegasemissions}}
  \label{fig:figure_co2emissions_map_2019}
  \end{figure}
  }

  \ifthenelse{\equal{\printtext}{true}}
  {
  
  Although the link between the rise in global temperature and the increase in extreme weather events has been scientifically proven, governments still have to contend with disbelief and lobbies that mislead measures to reduce local GHG emissions\cite{GlobalRegionalIncreasePrecipitationExtremesUnderGlobalWarming}.
  The impacts are already happening, in the form of extreme weather events  (EWE) \cite{CLARKE2021100285}, putting urban systems and infrastructure over eminent risk \cite{SANGSEFIDI2023104914}.
  
  According to \cite{EmissionsReportingReview}, the link between GHG emissions and economic activity is well established, as is the disconnect between environmental and social responsibility in measuring corporate performance. One of the reasons highlighted by the authors is the non-reporting of emissions, a recurring problem also seen in emissions reporting by local governments. 
  Emissions data are widely available from a variety of sources. The EPA maintains a catalogue of four climate change indicators (CCI) related to GHG emissions. Figure \ref{fig:figure_ghgs_emissions_histogram_example} shows the increase in GHG emissions from 1990 to 2015, but examples of emissions reporting that efficiently and effectively contribute to emissions reduction through mitigation actions are still hard to find.

  }

  \ifthenelse{\equal{\printfigures}{true}}
  {
  \begin{figure}[ht]
  \centering
  \includegraphics[width=\columnwidth]{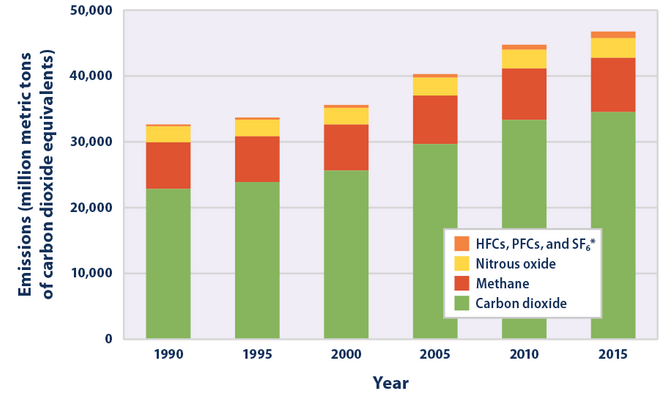}
  \caption[Global greenhouse gas emissions by gas, 1900-2015]{Global greenhouse gas emissions by gas, 1900-2015. Source: https://www.epa.gov/climate-indicators/climate-change-indicators-global-greenhouse-gas-emissions.}
  \label{fig:figure_ghgs_emissions_histogram_example}
  \end{figure}
  }
  
  \subsection{GHG impacts mitigation initiatives}

  \ifthenelse{\equal{\printnotes}{true}}
  {
  \begin{mdframed}[leftmargin=10pt,rightmargin=10pt]
  \color{red}
  \begin{itemize}
  \item CDP Initiative..
  \item GCOM Initiative and others..
  \item ? Cities are capable of emissions reduction policymaking based on available data ? 
  \end{itemize}
  \end{mdframed}
  }

  \ifthenelse{\equal{\printtext}{true}}
  {

  Cities are places of high overall primary energy consumption and high GHG emissions \cite{ghgandcities}\cite{urbanclimatechangemitigation}. The Carbon Disclosure Project (CDP)\footnote{https://www.cdp.net/en} is an initiative that promotes collaboration on emissions reduction and focuses on obtaining reliable data from cities and businesses worldwide to help them manage their environmental impacts. To drive the exploration and analysis of the data, CDP enlisted the infrastructure and expertise of Kaggle\footnote{https://www.kaggle.com/c/cdp-unlocking-climate-solutions} to promote a competition whose main objective was to discover key performance indicators (KPIs) among the responses provided. The database provided is based on questionnaires that CDP deployed in 2018, 2019 and 2020 to some cities and companies around the world. Cities are the ideal framework for implementing low-carbon policies \cite{acceleratinglowcarbondevelopment}. Also, the lack of Local Climate Plans(LCPs) is often connected to a lack of resources and capacity of local governments to tend to climate planning \cite{{CitiesMitigationPlans}}. It is also related to multi-level governance systems in which the upper levels of government do not set policy frameworks that encourage and guide local climate action \cite{GovernanceStructuresInClimateActions}. Thus, this work attempts to show the relationship between the information provided and the policies already in place that lead to emissions reduction and the associated benefits, both locally and globally. 
  
  Other initiatives, such as The Global Protocol for Community-Scale Greenhouse Gas Emissions Inventories guidelines, as pointed by \cite{greenhousegasemissionsinventory}, the Global Covenant of Mayors for Climate and Energy (GCoM)\footnote{https://www.globalcovenantofmayors.org/} and C40 cities\footnote{https://www.c40.org/}, are also the subject of this work, as they provide complementary and useful information on emissions at the city level. Despite the efforts of the selected cities, there are still some problems to be solved in emissions reporting in order for these cities to effectively contribute to the reduction of GHG emissions. Other approaches, such as \cite{DATOLA2023104821} can empower cities in the decision-making process.    

  }
  
  \subsection{Emissions reporting analysis using AI}

  \ifthenelse{\equal{\printnotes}{true}}
  {
  \begin{mdframed}[leftmargin=10pt,rightmargin=10pt]
  \color{red}
  \begin{itemize}
  \item Overview of Emissions Reporting Analysis process..
  \item Overview of AI techniques used in the process..
  \end{itemize}
  \end{mdframed}
  }
  
  \ifthenelse{\equal{\printtext}{true}}
  {

  Differences in emission levels depend on specific local features, such as climate conditions, urban form and density \cite{greenhousegasimplications}\cite{urbanco2mitigation}. Emissions reporting analysis can be made using statistical tools and techniques, also known as Analytics. This approach has already been used to produce relevant information in the field of emissions impact analysis \cite{AnalyticsCarbonEmissions}\cite{SustainabilityAccountingReporting}, such as indicators and correlations with external indices\cite{LowCarbonIndicator}, but it lacks a qualitative view of the data, which AI can also help with, and this is one of the analysis mechanisms used in this work.
  
  More than ever, algorithms and artificial intelligence techniques play a key role in every field of knowledge, especially when it comes to solving problems through optimization \cite{OptimizationUsingAITechniques}. Also, in the challenges and problems related to emissions reporting, these algorithms and techniques can be used to address and even solve some of them, such as data processing, integrity and usefulness. Those features represent the foundation for the development of policies based on predictions \cite{WheaterForecastingWithMLandHierarchicalClustering}. 
  
  }

  \subsection{Performance indicators for emissions}

  \ifthenelse{\equal{\printnotes}{true}}
  {
  \begin{mdframed}[leftmargin=10pt,rightmargin=10pt]
  \color{red}
  \begin{itemize}
  \item Overview of Performance Indicators development process..ok
  \item Overview of Performance Indicators applied to Emissions..ok
  \end{itemize}
  \end{mdframed}
  }

  \ifthenelse{\equal{\printtext}{true}}
  {

  Performance indicators (PIs) are one of the most commonly used tools for evaluating processes in terms of their effectiveness \cite{performancemeasurment} and their use is ubiquitous in the public sector \cite{performanceindicatorpublicsector}. Key Performance Indicators (KPIs) is the optimun result of PIs selection and is tightly correlated to data processing based on artificial intelligence methods and technologies, as pointed by \cite{BigDataWithMLandKPI}. Therefore, processes related to emissions reporting can also benefit from the concepts and formalization of the performance indicator development process. To achieve this goal, a performance indicator development process (PIDP) should be applied to emissions reporting processes to proceed with assessments based on available data \cite{publicsectordata}.
  
  The emissions reporting processes are subject to the PIDP, in which the analysis of candidate PI plays an important role, as this PI will be used to improve them. Thus, the candidates for PI can be used both to evaluate the effectiveness of the overall emissions reporting process and to search for other PIs candidates among the relationships with external indices and indicators.

  The PIDP depicts the basic concepts of PIs implementation. For this aspect, it can be seen as a framework to achieve the PIs implementation's goals, as a stakeholder-centred process developed by \cite{stackeholdercenteredprocess}

  }

  \color{black}

  \section{Performance Indicators Development Process}

  \ifthenelse{\equal{\printnotes}{true}}
  {
  \begin{mdframed}[leftmargin=1pt,rightmargin=1pt]
  \color{red}
  \begin{itemize}
  \item Performance Indicators Development Process definitions..
  \item List steps used to implement it..
  \subitem Data source selection..
  \subitem Data exploring..
  \subitem Data pre-processing..
  \subitem Quantitative Analysis..
  \subsubitem ClusWiSARD..
  \subsubitem Hierarchical Cluster..
  \subsubitem K-means..
  \subsubitem DBSCAN..
  \subsubitem Prevalence matrix..
  \subitem Qualitative Analysis..
  \subsubitem Grounded Theory..
  \subsubitem Case Study..
  \end{itemize}
  \end{mdframed}
  }

  \ifthenelse{\equal{\printtext}{true}}
  {

  The Performance Indicators Development Process (PIDP) explores and processes emissions-related data to look for candidates for performance indicators. The PIDP uses clustering techniques to group cities with similar answers to CDP forms questions. First, the data is downloaded from available sources of emissions-related data. The samples with standardized data are analyzed to search for performance indicators among the features representing these answers. The PIDP also deals with difficulties regarding the quality of the data being processed. The same process is presented by other techniques like data envelopment analysis (DEA), as pointed by \cite{DEAWithInperfectData}.
  
  \ifthenelse{\equal{\printfigures}{true}}
  {
  \begin{figure}[ht]
  \centering
  \includegraphics[width=\columnwidth]{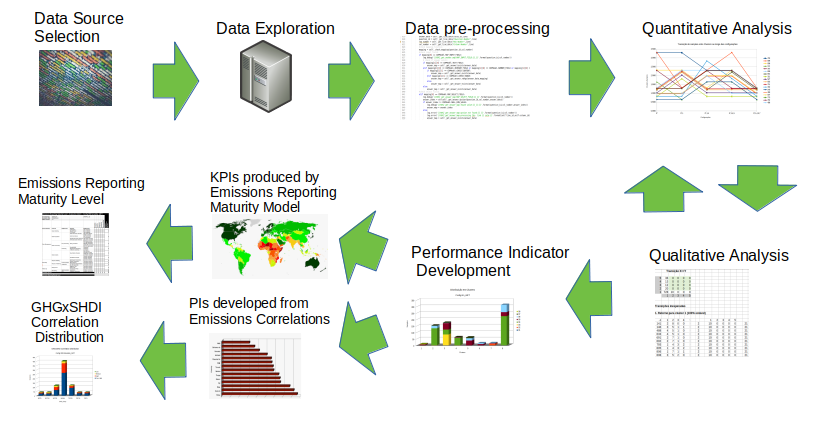}
  \caption[Performance indicators development process (PIDP) general view]{Performance indicators development process (PIDP) general view}
  \label{fig:figure_pidp_global_view}
  \end{figure}
  }
  
  The results of the clustering methods are used as the input to analyses to be done in the quantitative analysis phase. The quantitative analysis identifies potential performance indicator candidates among the questions used to segregate the samples into groups. 
  
  The confirmation of a candidate can be validated by looking into selected samples' data to check how stable is the candidate in separating the groups in the presence of other data. This step, executed using two experimental techniques, results in the qualitative analysis of the candidates to performance indicator. If the performance indicator found during the process can be used as the main source in a decision-making process, it is promoted to a key performance indicator. One example of this is shown in the emissions reporting maturity model (ERMM), a product of this process. The PIDP workflow overview is synthesized in Figure \ref{fig:figure_pidp_global_view}.

  }
  
  \subsection{Data sources selection and processing}

  \ifthenelse{\equal{\printnotes}{true}}
  {
  \begin{mdframed}[leftmargin=10pt,rightmargin=10pt]
  \color{red}
  \begin{itemize}
  \item General view..
  \item CDP <= Kaggle..
  \item CDP general view..
  \item Deliverables of this step..write
  \subitem ModelBase.py ..
  \subitem Databases in csv format..
  \end{itemize}
  \end{mdframed}
  }

  \ifthenelse{\equal{\printtext}{true}}
  {

  The first step in the performance indicators development process (PIDP) is to obtain reliable data about emissions among a set of cities representing as best as it can be the diversity found in the development level of cities along with the world. Thus, cities represent a minimal viable comparison unit in this work. The general view of this step is described in Figure \ref{fig:figure_pidp_datasourceselection_view}

  }
  
  \ifthenelse{\equal{\printfigures}{true}}
  {
  \begin{figure}[ht]
  \centering
  \includegraphics[width=\columnwidth]{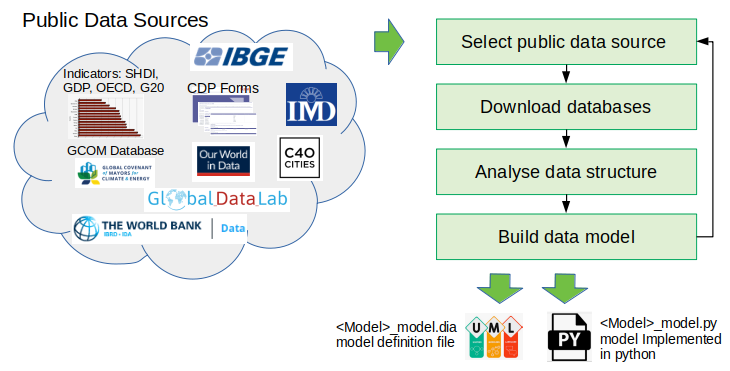}
  \caption[Data source selection general view]{Data source selection general view.}
  \label{fig:figure_pidp_datasourceselection_view}
  \end{figure}
  }

  \ifthenelse{\equal{\printtext}{true}}
  {
  
  According to the evaluation of the literature review focused on emissions reporting related topics \cite{EmissionsReportingReview}\cite{LowCarbonIndicator}\cite{SustainabilityIndicators}\cite{LCCIndicators} and data sources with emissions information from cities (GCoM\footnote{https://www.globalcovenantofmayors.org/our-cities/}, C40 cities\footnote{https://www.c40.org/cities}, Our World In Data (OWID)\footnote{https://github.com/owid/owid-datasets/tree/master/datasets}, Global Data Lab\footnote{https://globaldatalab.org}, World Bank\footnote{https://data.worldbank.org}, Instituto Brasileiro de Geografia e Estatística (IBGE)\footnote{https://cidades.ibge.gov.br/}), the CDP disclosure database demonstrated to be the most promising emissions reporting source of information. Thus, a public data source's list is built manually by analysing previous works, and it represents the main prerequisite to start the PIDP process. 
  
  CDP is an initiative to leverage policies and actions regarding reducing GHG emissions and their effects. It has registered 814 cities worldwide in 2019. This work uses the CDP database as the primary source of information, despite it having been retrieved from a secondary data set from a contest promoted by Kaggle. However, as pointed out by \cite{DataSetAndDataQuality}, the pattern of using secondary data, typically data sets that have been made publicly available through various repositories, remains the norm.
  
  }
  
  \ifthenelse{\equal{\printfigures}{true}}
  {
  \begin{figure}[ht]
  \centering
  \includegraphics[width=\columnwidth]{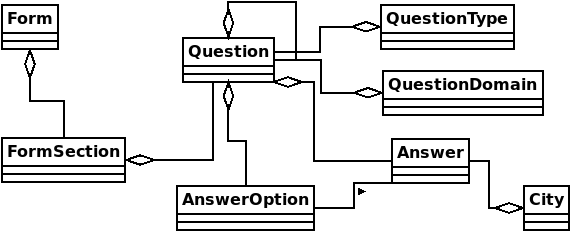}
  \caption[CDP UML model schema]{CDP UML model schema.}
  \label{fig:figure_pidp_cdp_model}
  \end{figure}
  }

  \ifthenelse{\equal{\printtext}{true}}
  {

  This first step in the PIDP process starts with obtaining the data and the underlying data structure from one of the data sources in the list. After downloading and checking the databases, the data structure is mapped to build a model to process the data. The model is built using unified modelling language (UML), proposed by \cite{UML}, for simplification and standardisation. Figure \ref{fig:figure_pidp_cdp_model} is shown as an example of a UML schema of the CDP model developed in this work. The process continues with the implementation of the UML model through a python file. It will be used in the data preprocessing phase and the clustering step of the quantitative analysis phase.   

  }

  \subsection{Data exploration}

  \ifthenelse{\equal{\printnotes}{true}}
  {
  \begin{mdframed}[leftmargin=10pt,rightmargin=10pt]
  \color{red}
  \begin{itemize}
  \item General view..
  \item CDP database detail..review
  \item Other databases..review
  \item Deliverables of this step..write
  \subitem cleaned and normalized data file..
  \subitem log file..
  \end{itemize}
  \end{mdframed}
  }

  \ifthenelse{\equal{\printtext}{true}}
  {
  
  The next step in the PIDP is to narrow the emissions-related data into data units. A data unit is an abstraction extracted from the data structure (answers field in the CDP database, for example) that can provide insights on candidates to performance indicators. This schema view of the data exploration step is shown in Figure \ref{fig:figure_pidp_dataexplporation_view}. 

  \ifthenelse{\equal{\printfigures}{true}}
  {
  \begin{figure}[ht]
  \centering
  \includegraphics[width=\columnwidth]{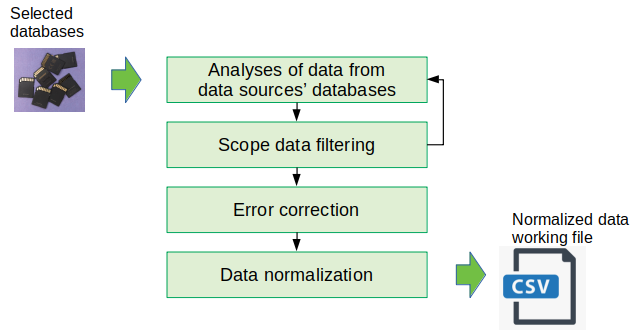}
  \caption[Data exploration general view]{Data exploration general view.}
  \label{fig:figure_pidp_dataexplporation_view}
  \end{figure}
  }
  
  The data to be explored should be initially reduced to represent a context (period of time, origin of the data, focus of the study - cities). Thus, the main goal of this scope filtering is to filter the valuable data among the data available from the databases. As an example, the CDP disclosure database has much information about emissions-related areas like transportation and energy, but the focus of this work is on emissions direct information, as presented in the sections described in table \ref{tab:cdp_sections}. In its third column (points) is represented the number of questions and sub-questions (tables) potentially used as the source of information. It indicates the potential of providing useful information in each form section.
  
  The process initiates attempts to correct errors such as inconsistencies found in the available data: e.g. wrong data type, empty value in "selection" or "multi-selection" answer type, empty value in "not null" answer. If the error cannot be recovered using other data from the same record, the record is discarded. 
  
  Other sources of emissions information were used to complete the information extracted from the CDP database. As an example, the GCoM congregates in a group of more than ten thousand cities. Its database was used to provide additional data about total emissions per year (2019), the presence of preparation (planning) to face emissions hazards and mitigation targets. The process uses other databases like gross domestic product (GDP), sub-national human development index (SHDI) and smart cities index (SCI).
    
  The data normalisation occurs when these additional data are joined to CDP data to produce useful information. Finally, external indicators such as GDP and SHDI are examples of this. The result of the processing is saved in a working file to be used as input in the data preprocessing phase.  
  
  \ifthenelse{\equal{\printnotes}{true}}
  {
  \begin{mdframed}[leftmargin=10pt,rightmargin=10pt]
  \color{red}
  \begin{itemize}
  \item Detail of questions and its organisation (base question, sub-questions, multi-value tables, etc
  \item Rules of answering not clearly follow
  \end{itemize}
  \end{mdframed}
  }

  \ifthenelse{\equal{\printtables}{true}}
  {
  
  \begin{table}[ht]
  \caption[CDP disclosure sections]{CDP disclosure sections. The column "points" represents how many questions and sub-questions could be used to retrieve useful information. \\ }
  \label{tab:cdp_sections}
  \centering
  {\footnotesize
  \begin{tabular}{p{2.5cm} p{3cm} l}
    \hline
    Section & Description & Points \\
    \hline
    0:Introduction & General information & 6 \\
    1:Governance and Data Management & Data management related information & 32 \\
    4:City-wide Emissions & Emissions produced by the city, its companies and citizens & 86 \\
    5:Emissions Reduction & Emissions reduction inventory reporting & 110\\
    7:Emissions Reduction by local government & Emissions reduction inventory of government scope & 40 \\
    \hline
  \end{tabular}}
  \end{table}
  }
  
  }
  
  \subsection{Data preprocessing}

  \ifthenelse{\equal{\printnotes}{true}}
  {
  \begin{mdframed}[leftmargin=10pt,rightmargin=10pt]
  \color{red}
  \begin{itemize}
  \item General view..
  \item Why the need for preprocessing..
  \item Filtering: questions, types of questions, samples
  \item Types of outputs RAW, BIN, DAT, CSV
  \item Statistics
  \item Scaling (log)
  \end{itemize}
  \end{mdframed}
  }
  
  \ifthenelse{\equal{\printtext}{true}}
  {

  The data preprocessing step is responsible for preparing the available data to be correctly used by the clustering algorithms and it requires the data to be cleaned from consistency errors. After an initial inspection of the data in the CDP database forms, inconsistencies and errors were found that could jeopardise the clustering process. 
  
  \ifthenelse{\equal{\printfigures}{true}}
  {
  \begin{figure}[ht]
  \centering
  \includegraphics[width=\columnwidth]{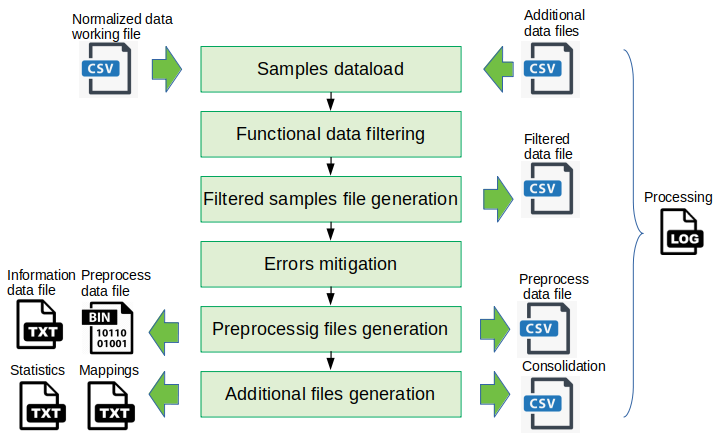}
  \caption[Data preprocessing view]{Data preprocessing view.}
  \label{fig:figure_pidp_datapreprocessing_view}
  \end{figure}
  }
    
  According to \cite{DataQualityNASA}, the best preprocessing strategy is that in which the problematic data should be treated first. Some cases of either conflicting feature values or implausible values should be discarded before data can be used. Therefore, it was necessary to build a support system to deal with these issues and leverage the quantitative and qualitative analysis steps. Figure \ref{fig:figure_pidp_datapreprocessing_view} shows the data preprocessing schema with the generated output files. The generation of the files, their usages and which goals they address are detailed in the following sections.

  \textbf{Input data}

  A working file provides the input normalised data and additional data files in CSV format. Each one of the composed databases has a related model to support the processing of the underlying information. The models define the fields, the fields' types, and the operations realised over the data. For example, for fields of type "single selection" and "multiselection", the models also check the values provided. For example, in fields with types "date" and "year": the ranges are defined to help validate the values informed. 
  
  For the CDP database, the classes that implement the concepts of fields, types, domains and operations are listed in table \ref{tab:cdp_database_mapping}.

  \ifthenelse{\equal{\printtables}{true}}
  {
  \begin{table}[ht]
  \caption[CDP database classes mapping]{CDP database classes mapping. The mappings show the relation between the CDP data model and data processing concepts. \\ }
  \label{tab:cdp_database_mapping}
  \centering
  {\footnotesize
  \begin{tabular}{p{2cm} p{5cm}}
    \hline
    Class & Mapped concept \\
    \hline
    Form & Data set \\
    FormSection & Sub data set \\
    Question & Field \\ 
    QuestionType & Field type: null, not null, single-select, multi-select \\
    QuestionDomain & Field type: DATE, YEAR, NUMBER, INTEGER, TEXT \\
    Answer & Field values \\ 
    AnswerOption & Field values options; case of single-select of multi-select types \\
    City & Record \\
    \hline
  \end{tabular}}
  \end{table}
  }

  \textbf{Samples dataload}
  
  The CDP disclosures are organised in forms, sections, questions, sub-questions and options used as answers. The CDP model maps the questions and answers' options into fields that can be used as features along the process. Each question and sub-question is represented by a line in the input data file. The table \ref{tab:input_csv_fields} details the cells presented in the line to process. Thus, the data load of the samples is the first step inside data preprocessing, in which the normalised data working file is loaded along with the additional information present in additional data files.

  \ifthenelse{\equal{\printtables}{true}}
  {
  \begin{table*}[ht]
  \caption[Structure of a line in the input data file]{Structure of a line in the input data file from CDP database in CSV format. \\ }
  \label{tab:input_csv_fields}
  \centering
  {\footnotesize
  \begin{tabular}{p{3cm} p{5cm} p{7cm}}
    \hline
    Cell & Description & Remarks \\
    \hline
    Questionnaire & Form identification & Filtered: Cities 2019 \\
    Year Reported to CDP & Base year for answers in & Filtered: 2019 CDP database \\
    Account Number & Unique identification for city in CDP database (Sample Id) & Unique Id \\ 
    Organisation & City Name & Normalized to include State Name for clarification \\
    Country	CDP Region & CDP regions defined in table \ref{tab:cdp_regions_list} & \\
    Parent Section & Group of sections & \\
    Section & Group of answers & \\
    Question Number & Question unique identification & \\
    Question Name & Question unique name & \\
    Column Number & Column unique identification inside question & Column 0 indicates direct answer \\
    Column Name & Column name to identify tabled answer & \\
    Row Number & Row unique identification inside question & Row 0 indicates direct answer \\
    Row Name & Row name to identify tabled answer & \\
    Response Answer & Answer value & \\
    Comments & & Used to clarify the answer \\
    File Name & & Complementary information about external file \\
    Last update & Data time of last update of the record & \\
    \hline
  \end{tabular}}
  \end{table*}
  }
  
  \textbf{Functional data filtering}
  
  Functional data filtering occurs when filtering parameters are passed to preprocessing execution module to segregate only the information needed in the context of a preprocessing configuration and optimise the drill down during quantitative and qualitative analyses. To support these analyses, this step generates a "filtered data file", which is an exact copy of the filtered samples. The filtering engine can be used to select a set of questions and sub-questions, a set of samples (listed using a samples file) or all samples in which a field type is present. The filtering engine permits include (I:) or exclude (E:) operators, acting to compose the filtering rules to be applied to the data. Some filtering examples are shown in table \ref{tab:filtering_examples}. 
 
  \ifthenelse{\equal{\printtables}{true}}
  {
  \begin{table*}[ht]
  \caption[Filtering examples]{Filtering examples used during the experimental phase of this work. \\ }
  \label{tab:filtering_examples}
  \centering
  {\footnotesize
  \begin{tabular}{p{10cm} l}
    \hline
    Filtering scope & Filtering options \\
    \hline
    Question "0" and its sub-questions & I:Question\&nbsp;Number=0* \\
    All questions "0,1,4,5,7" and their sub-questions & I:Question\&nbsp;Number=0*,1*,4*,5*,7* \\
    All questions "0,1" and their sub-questions, excluding fields with type YN and Number=0*,1* & E:\#FieldType=YN;I:Question\&nbsp; \\
    All questions reported by cities in samples.txt file & "I:\#SampleId=@samples.txt \\
    \hline
  \end{tabular}}
  \end{table*}
  }
  
  \textbf{Errors mitigation}

  One problem identified in the CDP forms data entry is the text representation for questions with single and multi-selection options. To solve this issue, the CDP model implements unique codes and associates them with the available options. However, in some samples, the text informed does not match the text of any option available to that question. In this situation, the use of techniques to correct the string representation based on the number of changes, like Damerau-Levenshtein distance, as presented by \cite{StringCorrectionUsingDL}. The table \ref{tab:DL_answers_correction} shows some examples found in CDP database preprocessing. Thus, the need for an "errors mitigation" step in data preprocessing is to leverage the use of defective data in the subsequent phases. The techniques applied in error correction depend on the nature of the error: e.g. domain-value matching, invalid value type, and values out-of-ranges.

  \ifthenelse{\equal{\printtables}{true}}
  {
  \begin{table*}[ht]
  \caption[Examples of application of Damerau-Levenshtein distance to answers correction]{Examples of application of Damerau-Levenshtein distance to answers correction. The text differences are presented in bold. \\ }
  \label{tab:DL_answers_correction}
  \centering
  {\footnotesize
  \begin{tabular}{p{1cm} p{1cm} p{6cm} p{3cm} p{3cm}}
    \hline
    CDP Id & City Name & Question & Original Answer & Correct Answer \\
    \hline
    1093 & Atlanta & 1.1a: Please select any commitments to climate adaptation and/or mitigation your city has signed and attach evidence & Individual city \textbf{c}ommitment & Individual city \textbf{C}ommitment \\
    \hline
    1184 & Austin & 1.13:What tools does your city/department use to analyse its environmental-related data? Select all that apply. & Visualization/Analysis Software - Tableau \textbf{;} Qlik etc & Visualization/Analysis Software - Tableau \textbf{,} Qlik\textbf{,} etc \\
    \hline
    1184 & Austin & 5.0a: Please provide details of your total city-wide base year emissions reduction (absolute) target. & Larger \textbf{–} covers the whole city and adjoining areas & Larger \textbf{-} covers the whole city and adjoining areas \\
    \hline
  \end{tabular}}
  \end{table*}
  }

  The invalid value type occurs when a numeric value is expected, and a "null" or other value type is provided to an answer instead. The mitigation, in this case, is to convert the text representation to the best number representation, when it is possible, and to set the answer to "zero value" and "not answered" when it is not. The values out-of-range issue is mitigated using statistics tools (e.g.variance) to check and correct scaling errors. To achieve this, the model used to support the processing of the database holds the expected min and max (range) values that are supposed to happen and a "mark" in the question in the model indicating that it has to be range-checked. 

  \textbf{Preprocessing files generation}

  The internal representation of the files differs based on the target clustering engine that will be used. The textual representation will be used as the input file for Hierarchical, K-means and DBSCAN clustering methods. The file with binary representation, on the other hand, will be used as input for ClusWiSARD. 
  
  \ifthenelse{\equal{\printtables}{true}}
  {
  \begin{table*}[ht]
  \caption[Conversion mechanisms used to transform the input CSV file into processed textual representation file]{Conversion mechanisms used to transform the input CSV file into processed textual representation file, also in CSV format. The result of the data processing is saved in the correspondent textual processed data file. \\ }
  \label{tab:raw_to_pre_conversion_mechanisms}
  \centering
  {\footnotesize
  \begin{tabular}{p{2cm} p{14cm}}
    \hline
    Field type & Conversion mechanism \\
    \hline
    TABLE & Conversion of each field's value inside the table (multi-select) using the correspondent field type rule described here. Each value is separated by ":" in a list representing each row of the table of multi-select fields. \\
    SELECT & Conversion to the numeric value represented the text informed in the field value. If the field value is not found among the predefined answering options, error mitigation techniques try to choose the best available option. If it is not possible, the conversion uses "0" to represent the "not found" answering option. \\    
    TEXT & Conversion to "0" if the field is empty or "1" on the contrary. \\
    NUMBER & Conversion to the log of the field's value to try to narrow to a common scale to be used with the other questions. The log value is then converted to a text representation. \\
    INTEGER & Conversion straight to text representation as-is. \\
    YEAR & Conversion of difference from base year value to text. \\
    DATE & Conversion to ISO data format (ISO 8061) without hyphenation. \\
    \hline
  \end{tabular}}
  \end{table*}
  }

  During processing, each field generates an output in text format, based on the rules defined by the model. The field type and specification define the field's value conversion mechanism. The table \ref{tab:raw_to_pre_conversion_mechanisms} details the conversion mechanisms used. Due to optimisation, during the preprocessing of the numeric fields (NUMBER, INTEGER and YEAR types), some statistics are collected to be used in the next binarisation step. To obtain the best minimum value for the number of bits, the data preprocessing uses the number of options for an answer, the single and multi-selection fields, and the number of digits in the answer for the numeric fields. The number obtained is registered as the \textbf{binary slot size} in the information data file generated by the process. It is used to define the same number of bits applied to all answers.
  
  \ifthenelse{\equal{\printtables}{true}}
  {
  \begin{table*}[ht]
  \caption[Conversion mechanisms used to transform the processed textual representation file in CSV format into processed binary representation file]{Conversion mechanisms used to transform the processed textual representation file in CSV format into processed binary representation file, also in CSV format. The result of the data processing is saved in the correspondent binary representation processed data file.  \\ }
  \label{tab:pre_to_bin_conversion_mechanisms}
  \centering
  {\footnotesize
  \begin{tabular}{p{2cm} p{14cm}}
    \hline
    Field type & Conversion mechanism \\
    \hline
    TABLE & Conversion is applied to each value in the list of preprocessed text values according to the rules described here. The final binary value is a superposition ("OR" operation) of each bit of each binary value of each field in the table. \\
    SELECT & Conversion to bits-value representation using two techniques: bit-mapping and thermometers. The bit-mapping is used for multi-select fields and maps the numeric value of the option chosen as an index to the position in the bit string, which is filled by $s$-bits "1". The number of $s$-bits is a result of the \textit{slot size} divided by the total number of options for the answer. The thermometer technique adds "1"s bits to fill the string (from left to right) until reached the position of the option. This technique is used when processing single-select fields. \\
    TEXT & Conversion to full "0"s or "1"s depending on the processed value. \\
    NUMBER & Conversion to the thermometer representation of the processed value. The mechanism is the same as the one applied to SELECT field, but using min and max values computed along the answers to establish the \textit{scale} of the thermometer. Thus, the number of bits used is the result of slot size times the field value minus the min value divided by the max value minus the min value.  \\   
    INTEGER & Conversion to thermometer representation as described in NUMBER field. \\
    YEAR & Conversion to thermometer representation as described in NUMBER field. \\
    DATE & Conversion to thermometer representation as described in NUMBER field. \\
    \hline
  \end{tabular}}
  \end{table*}
  }
  
  To achieve the best comparison results from the AI engines used in this work, it is necessary to guarantee that different clustering methods use the same clustering information in different formats. During the generation of the binary file, the text values are converted into binary (0's and 1's) representation, based on the field type and specification. Hereafter, the binarization step occurs when the binary representation file is generated based on another conversion mechanism applied over the processed textual representation file. The table \ref{tab:pre_to_bin_conversion_mechanisms} details the conversion mechanisms used to generate the binary representation of the processed textual data. To avoid misinterpretation of which file should be used as input to ClusWiSARD, the file with binary representation content receives a \textit{.bin} extension.
    
  An example depicting the data processing of the city of Rio de Janeiro's data extracted from CDP forms, present in the CDP forms database file, is shown in Figure \ref{fig:figure_pidp_datapreprocessing_rio_example}.
  
  \ifthenelse{\equal{\printfigures}{true}}
  {
  \begin{figure*}[ht]
  \centering
  \includegraphics[width=\linewidth]{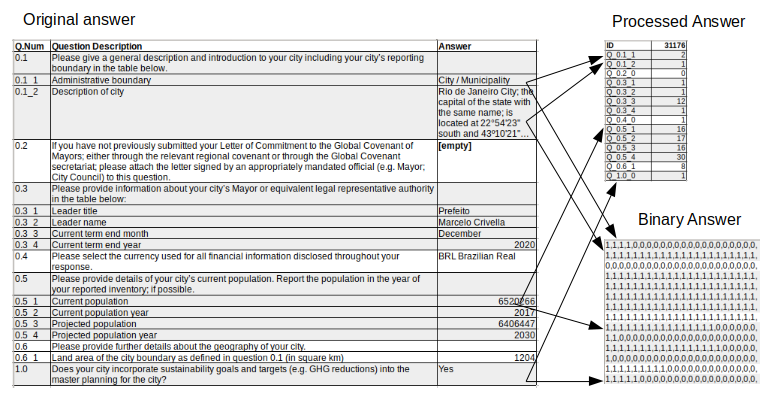}
  \caption[Preprocessing generation using Rio de Janeiro (31176)]{Preprocessing generation using Rio de Janeiro (31176) data sample example.}
  \label{fig:figure_pidp_datapreprocessing_rio_example}
  \end{figure*}
  }
  
  \textbf{Additional files generation}

  During preprocessing, some additional files are generated as important byproducts. The consolidation data file holds information about the processed numeric values: min, max, mean, and frequency of not empty answers. These values make it possible to check the distribution behaviour observed using the thermometer technique to process binary data output.
  
  The questions filtered in the preprocessing are put in a list with question\_id and question\_name. At the end of the preprocessing step, a text file is saved with the number and description of the question. It is used to facilitate the qualitative analysis based on the applicability of the questions. For questions in which the underlying field of type is multi-select, the options are also listed to help calibrate the quantitative analysis as needed. The questions of configuration 0a1a4a5a are listed in the appendix of this work.

  The processing statistics output file holds quantitative and qualitative information about the processing of questions for each city. Table \ref{tab:preprocessing_statistics} shows the details of the obtained statistics.  
  
  \ifthenelse{\equal{\printtables}{true}}
  {
  \begin{table}[ht]
  \caption[Preprocessing statistics details]{Preprocessing statistics details collected during the execution of the experiments. \\ }
  \label{tab:preprocessing_statistics}
  \centering
  {\footnotesize
  \begin{tabular}{|l|l|c|}
    \hline
    Statistic & Detail \\
    \hline
    SampleId & City unique identification \\
    TABLE & Count of fields of type "table" with answer \\
    SELECT & Count of multi-select field with answer \\
    TEXT & Count of fields of type "text" with answer \\
    NUMBER & Count of fields of type "number" with answer \\
    INTEGER & Count of fields of type "integer" with answer \\
    YEAR & Count of fields of type "year" with answer \\
    DATE & Count of fields of type "date" with answer \\
    CC\_R & Count of characters in the answer \\
    CC\_C & Count of characters in the comments \\
    WC\_R & Count of words in the answer \\
    WC\_C & Count of words in the comments \\
    WU\_R & Count of unique words in the answer \\
    WU\_C & Count of unique words in the comments \\
    WD\_R & Count of dictionary words in the answer \\
    WD\_C & Count of dictionary words in the comments \\
    \hline
  \end{tabular}}
  \end{table}
  }
  
  One extraction to exemplify the statistics obtained during preprocessing is shown in Figure \ref{fig:figure_pidp_datapreprocessing_stats_example}. The differences between the cities are established, even being part of the same south-east region. For example, despite having the best GDP, São Paulo is far from being the best information provider. 
  \ifthenelse{\equal{\printfigures}{true}}
  {
  \begin{figure*}[ht]
  \centering
  \includegraphics[width=\linewidth]{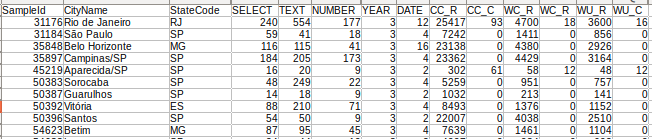}
  \caption[Preprocessing statistics extraction example]{Preprocessing statistics extraction example listing ten cities in Brazil.}
  \label{fig:figure_pidp_datapreprocessing_stats_example}
  \end{figure*}
  }
  
  \textbf{ Processing logs}

  The logging information generated during the preprocessing step is used to check the overall process and validate the information's reliability. The indication of errors in the logs interrupts the (next) output generation step, forcing checking what is causing it. For example, the CDP database has some errors in field mapping, domain values, and rules applied to form filling. These errors were marked or fixed to continue the form processing. Another use for general logging is to set up the proper provisioning for machine power and memory needed in preprocessing and the following steps. A logging extraction of the preprocessing phase is listed in the appendix of this work.

  }
  
  \section{Quantitative analysis}

  \ifthenelse{\equal{\printnotes}{true}}
  {
  \begin{mdframed}[leftmargin=10pt,rightmargin=10pt]
  \color{red}
  \begin{itemize}
  \item General view
  \item Clustering to group together similar answers to the CDP forms
  \item Compare results with other clustering methods
  \item Labeling using external indicators
  \end{itemize}
  \end{mdframed}
  }

  \ifthenelse{\equal{\printtext}{true}}
  {
  
  The quantitative analysis of the results obtained from the clustering methods can be used to indicate features with better chances to be used as performance indicators. Thus, the clustering results are treated and viewed as an alternative to purely statistical ones. However, the main goal is to search for similarities and answers that indicate different approaches implemented by the cities that are grouped in the same cluster. The nuances of the clustering process, the comparative data generated, and the validation techniques are shown in the following sections.

  Figures \ref{fig:figure_pidp_quantitativeanalysis1_view} and  \ref{fig:figure_pidp_quantitativeanalysis2_view} show a general view of this step of the process.

  \ifthenelse{\equal{\printfigures}{true}}
  {
  \begin{figure}[ht]
  \centering
  \includegraphics[width=\columnwidth]{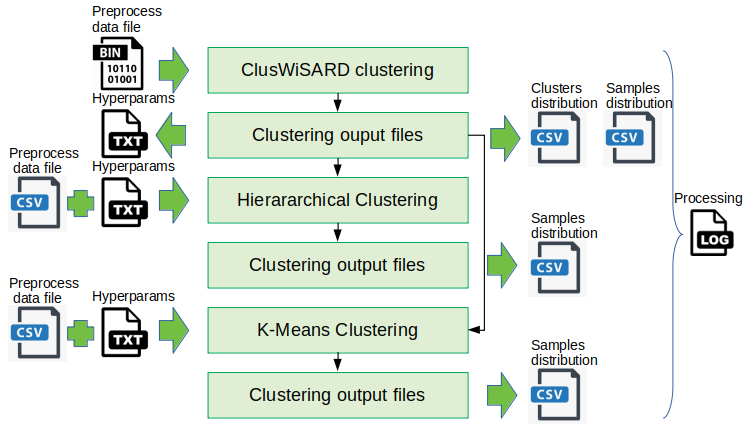}
  \caption[Quantitative analysis schema view]{Quantitative analysis schema view}
  \label{fig:figure_pidp_quantitativeanalysis1_view}
  \end{figure}
  }
  
  \ifthenelse{\equal{\printfigures}{true}}
  {
  \begin{figure}[ht]
  \centering
  \includegraphics[width=\columnwidth]{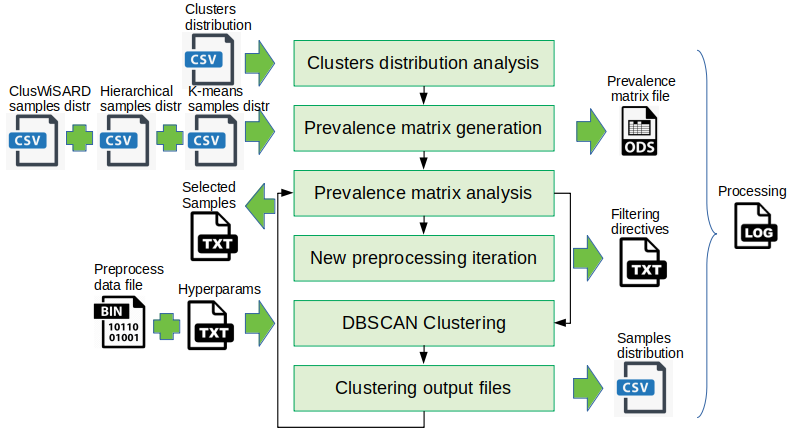}
  \caption[Quantitative analysis schema view (continuation)]{Quantitative analysis schema view (continuation)}
  \label{fig:figure_pidp_quantitativeanalysis2_view}
  \end{figure}
  }
  
  }
  
  \subsection{Using ClusWiSARD}

  \ifthenelse{\equal{\printnotes}{true}}
  {
  \begin{mdframed}[leftmargin=10pt,rightmargin=10pt]
  \color{red}
  \begin{itemize}
  \item General view
  \item Hyperparameters to use
  \end{itemize}
  \end{mdframed}
  }

  \ifthenelse{\equal{\printtext}{true}}
  {
  
  ClusWiSARD is the primary clustering mechanism used to group the samples (cities) with similar or related answers. The other clustering mechanisms were used to validate and narrow the quantitative analysis process in pursuing performance indicators based on the answers. The ClusWiSARD results can be seen as "pictures" taken from the binary correspondence of the CDP forms' responses and additional data. The similarities in the answers are registered and used to group the samples into clusters. 

  \ifthenelse{\equal{\printfigures}{true}}
  {
  \begin{figure}[ht]
  \centering
  \includegraphics[width=\columnwidth]{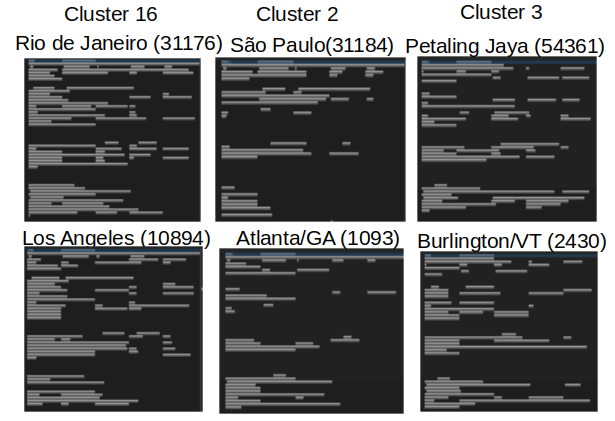}
  \caption[Preprocessing binary representation extraction example]{Preprocessing binary representation extraction example as "pictures". }
  \label{fig:figure_pidp_datapreprocessing_bin_extraction}
  \end{figure}
  }
  
  The main advantage of ClusWiSARD application to this work is the generalisation capacity of the method. As expected, the cities can provide different answers to the same question. But these differences in the pictures can be more subtle, making it harder to extract a pattern among them. Even though different, the answers set tends to generate patterns in the responses used to identify candidates to performance indicators. Figure \ref{fig:figure_pidp_datapreprocessing_bin_extraction} shows an example of "pictures" processed by ClusWiSARD. 
  
  This step is the generation of two CSV files: a cluster distribution and the distribution of a sample. The clusters distribution file holds information about how the clusters were consolidated. The number of clusters in which a sample can be grouped is registered along with the cluster chosen as the best choice (group) for this sample. This measures how stable is the clustering process given the hyperparameters informed by the ClusWiSARD algorithm. Figure \ref{fig:figure_cluswisard_clusters_distrib_example} shows clusters and samples distributions examples.
  
  \ifthenelse{\equal{\printfigures}{true}}
  {
  \begin{figure*}[ht]
  \centering
  \includegraphics[width=\linewidth]{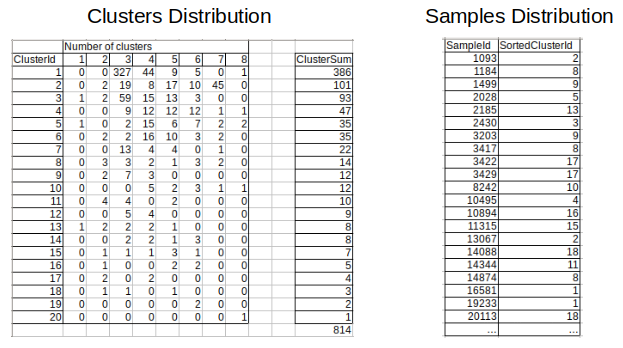}
  \caption[Clusters and samples distributions examples]{Clusters and samples distributions examples.}
  \label{fig:figure_cluswisard_clusters_distrib_example}
  \end{figure*}
  }
  
  To achieve better results, the ClusWiSARD is executed first  in "discover" mode as the hyperparameters threshold and discriminator\_limit are set to "auto" value. In this step, a text file with the best values for these two hyperparameters and the other hyperparameters used to execute the method is saved from being used in another process of hierarchical and k-means clustering methods.  
  
  }

  \subsection{Using other clustering methods}

  \ifthenelse{\equal{\printnotes}{true}}
  {
  \begin{mdframed}[leftmargin=10pt,rightmargin=10pt]
  \color{red}
  \begin{itemize}
  \item General view
  \item Describe other clustering methods' advantages 
  \item Hierarchical Clustering
  \item K-means
  \item DBSCAN
  \end{itemize}
  \end{mdframed}
  }

  \ifthenelse{\equal{\printtext}{true}}
  {
  
  Some other clustering methods were used in this work to validate the results of ClusWiSARD regarding the processing of the available data, as these other methods use different approaches to identify the groups of data (clusters). Among differences in implementation, ClusWiSARD uses a non-deterministic approach to group similar "pictures" from the data, as the other methods use a deterministic one. The Hierarchical Clustering method uses the aggregation (agglomerative or bottom-up approach) of similar features of the samples to compose the groups. The maximum number of groups (clusters) is pre-defined, and it is set as the same as the one used in ClusWiSARD. In addition to it, K-means uses another approach that uses the Euclidean distance between the field values to $k$ centroids (or geometric centres) to group the samples. Both methods have the results compared to decide the use (or not) of the DBSCAN method to complete the analysis.

  The execution of hierarchical and k-mean clustering methods uses the processed CSV format data file as input and the hyperparameters used in the ClusWiSARD method. The byproducts of this step are the files with the distributions of the samples that will be used to compose the prevalence matrix in the further step of the process.  
  
  The result of this step is the generation of the prevalence matrix file, as shown in Figure \ref{fig:figure_prevalence_matrix_example}. The prevalence matrix analysis leads to four possible paths:
  
  \begin{itemize}
  \item a new preprocessing iteration with a new configuration: when the analysis of the prevalence matrix indicates a "dead-end", a new filtering configuration is established and the preprocessing phase is executed again.
  \item a new preprocessing iteration with new filters and the selected samples generated by the prevalence matrix analysis: this path is based on the "drill-down" of the analysis of the set of samples that can hold information to lead to identifying performance indicators candidates, but still have to be verified through another iteration of the quantitative analysis so far.
  \item a selected sample set that will be analyzed in the qualitative analysis step: this path occurs when the analysis of the prevalence matrix indicates that the configuration being evaluated has a good chance to produce a performance indicator candidate. In this case, a selected sample file is generated to be used in the qualitative analysis phase.    
  \item a DBSCAN clustering execution with the same hyperparameters as the previous methods: this happens when the analysis of the relations between ClusWiSARD, hierarchical clustering and k-Means clustering did not point to a clear result. In this case, a DBSCAN method is executed to help identify a more clear path reducing the plausible "noise" in the samples analyzed so far in this step.
  \end{itemize}
  
  \ifthenelse{\equal{\printfigures}{true}}
  {
  \begin{figure*}[ht]
  \centering
  \includegraphics[width=\linewidth]{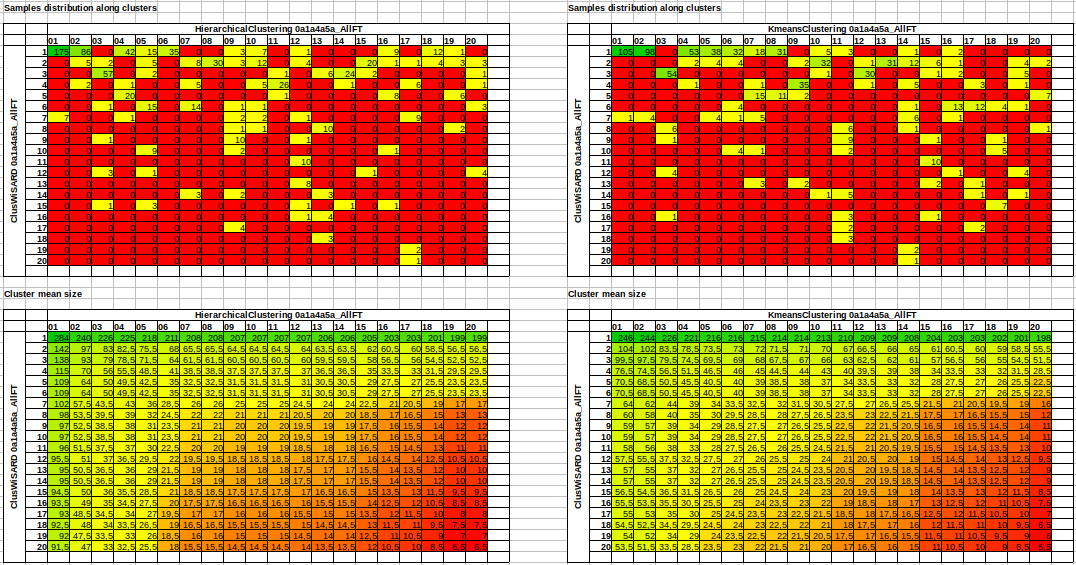}
  \caption[Prevalence matrix example]{Prevalence matrix example.}
  \label{fig:figure_prevalence_matrix_example}
  \end{figure*}
  }
 
  Thus, the validation step in the process is based on comparing the behaviour of ClusWiSARD with the other clustering methods. The samples in each cluster should be compared to their corresponding in the other clustering methods, generating a prevalence matrix $P$. 
  
  This matrix is built using the formula \ref{eq:prevalence_matrix}.\\
  
  \begin{equation}
  \label{eq:prevalence_matrix}
  P_{i,j} = ( 2 * b_{i,j} )  / ( c_{i} + v_{j} )
  \end{equation}
  where $P_{i,j}$ is the prevalence index in ${i,j}$;  $ i \in C $ ; $ j \in V $; $C$ is the ClusWiSARD clusters set; $V$ is the validation clustering mechanism (hierarchical clustering, k-means or DBSCAN) clusters set; $b_{i,j}$ is the number of samples present both in $ C_{i}$ and $V_{j}$; $C_{i}$ is a subset of C with samples in cluster $i$; $V_{j}$ is a subset of V with samples in cluster $j$; $c_{i}$ is the number of samples in $C_{i}$; $v_{j}$ is the number of samples in $V_{j}$.

  The samples present in the clusters with a higher prevalence index in $P$ are then selected, and another experiment is executed using the same hyperparameters as the original experiment. This analysis and verification processes repeat as long as the mean global prevalence index ($mpi$) is greater or equal to the prevalence index of the last experiment. 
  
  The $mpi$ is built using the formula \ref{eq:mpi_value}.\\
  
  \begin{equation}
  \label{eq:mpi_value}
  mpi = \sum_{\overset{1<i<m}{1<j<n}} P_{i,j} / i*j
  \end{equation}  
  where $mpi$ is the mean prevalence index; $P_{i,j}$ is the prevalence index in ${i,j}$ ; $m$ is the number of clusters identified by ClusWiSARD; $n$ is the number of clusters identified by the validation clustering mechanism (hierarchical clustering, k-means or DBSCAN).

  Other techniques, as pointed by \cite{AnalyticHierachyProcessWithMatrices}, follow the same path in evaluating relationships based on matrices representations.
 
  }

  \section{Qualitative analysis}

  \ifthenelse{\equal{\printnotes}{true}}
  {
  \begin{mdframed}[leftmargin=10pt,rightmargin=10pt]
  \color{red}
  \begin{itemize}
  \item General view
  \end{itemize}
  \end{mdframed}
  }

  \ifthenelse{\equal{\printtext}{true}}
  {
  
   The main objective of the qualitative analysis phase is to compare the responses of different cities present in the same cluster, indicating a convergent approach over the data. Another strategy is to compare different responses from cities in different clusters, indicating a divergent approach in this case. The following sections detail the techniques involved in the step of the process. 

  }
  
  \subsection{Using Grounded Theory}

  \ifthenelse{\equal{\printnotes}{true}}
  {
  \begin{mdframed}[leftmargin=10pt,rightmargin=10pt]
  \color{red}
  \begin{itemize}
  \item General view
  \item Describe adaptations needed to implement GT over data
  \end{itemize}
  \end{mdframed}
  }

  \ifthenelse{\equal{\printtext}{true}}
  {

  \ifthenelse{\equal{\printfigures}{true}}
  {
  \begin{figure}[ht]
  \centering
  \includegraphics[width=\columnwidth]{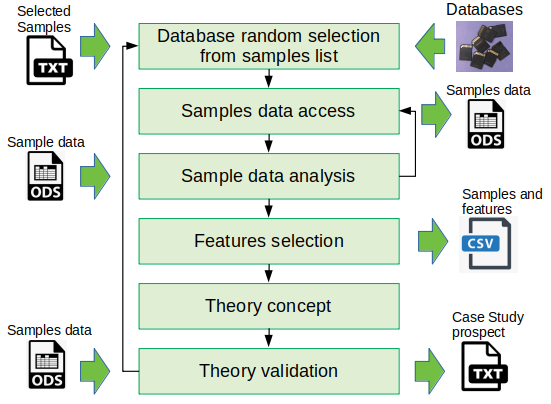}
  \caption[Qualitative Analysis: Grounded Theory]{Qualitative Analysis: Grounded Theory application general view.}
  \label{fig:figure_pidp_qualitative_groundedtheory_view}
  \end{figure}
  }
  
  This work uses an adaptation of what is proposed by Grounded Theory to facilitate analysing responses to the same question from different cities. Here, a random set of cities is gathered from the cities in a cluster, which was selected as the most promising from the quantitative analysis step. A set of questions of interest is chosen, and a matrix is built to allow visual analysis. According to the results, another round is performed to select other cities for comparison. This procedure is performed when the results are inconclusive or show a possible tendency in the answering process. This tendency composes a theory of answering that should be confirmed or denied in further steps. The next set of cities can be used to confirm the tendency, reinforce the theory, or deny it, resetting the process to look for another theory based on other tendencies. The analysis continues until more than 50\% of the cities are selected. Hereafter, if the tendency pattern remains, the process involves finding samples that represent exceptions to the theory (or candidate rule), using the subsequent (case study) approach.

  }
  
  \subsection{Using Case Study}

  \ifthenelse{\equal{\printnotes}{true}}
  {
  \begin{mdframed}[leftmargin=10pt,rightmargin=10pt]
  \color{red}
  \begin{itemize}
  \item General view
  \item Describe Case Study confirm/deny hypothesis strategy
  \item When we need to start over analysis or step back to quantitative
  \end{itemize}
  \end{mdframed}
  }

  \ifthenelse{\equal{\printtext}{true}}
  {

  \ifthenelse{\equal{\printfigures}{true}}
  {
  \begin{figure}[ht]
  \centering
  \includegraphics[width=\columnwidth]{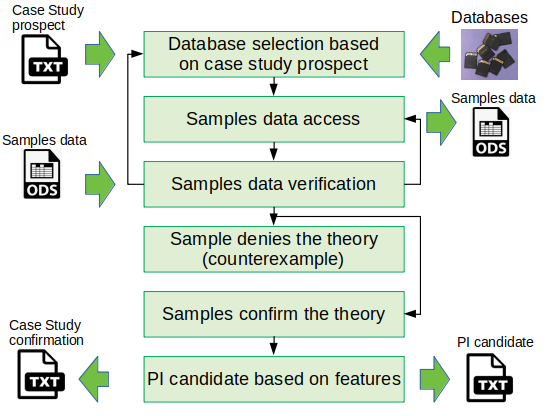}
  \caption[Qualitative analysis: Case Study]{Qualitative analysis: Case Study application general view.}
  \label{fig:figure_pidp_qualitative_casestudy_view}
  \end{figure}
  }
  
  The case study approach uses all available questions from a single city selected from any other than the selected cluster being analysed to check for inconsistencies that confirm or discredit a tendency found through the grounded theory approach. Suppose it is impossible to proceed with the confirmation or denial of the theory. In that case, another city is selected from another cluster, and the analysis continues until all clusters have been visited at least once.  
  
  }
  
  \section{Emissions Reporting Maturity Model}

  \ifthenelse{\equal{\printnotes}{true}}
  {
  \begin{mdframed}[leftmargin=1pt,rightmargin=1pt]
  \color{red}
  \begin{itemize}
  \subitem Overview of CMM and DMM..
  \subitem Emissions Reporting Maturity Model definitions..
  \subsubitem How ERMM uses PIDP
  \subitem Emissions Reporting Maturity Level definitions..
  \subsubitem How derivation of Emissions Reporting Maturity Level from ERMM takes place..
  \subsubitem How Emissions Reporting Maturity Level is affected by regions-related data..  \end{itemize}
  \end{mdframed}
  }

  \ifthenelse{\equal{\printtext}{true}}
  {

  \ifthenelse{\equal{\printfigures}{true}}
  {
  \begin{figure}[ht]
  \centering
  \includegraphics[width=\columnwidth]{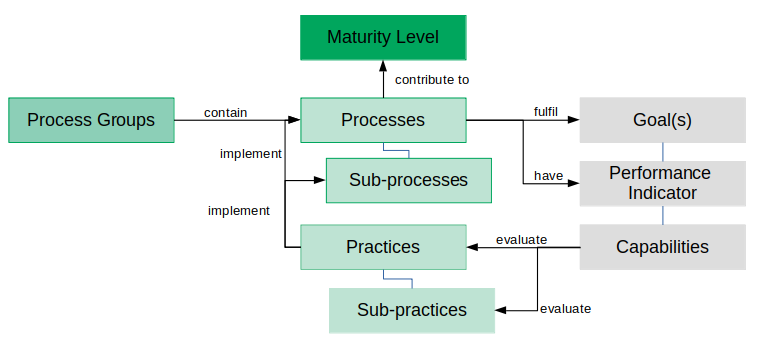}
  \caption[Emissions reporting maturity model general view]{Emissions reporting maturity model general view.}
  \label{fig:figure_ermm_general_view}
  \end{figure}
  }

  \begin{mdframed}[leftmargin=1pt,rightmargin=1pt]
  The emissions reporting maturity model (ERMM) stands for a methodology to select, process, classify and deliver evaluations of emissions-related processes based on the information presented in emissions reports.
  \end{mdframed}
  
  The ISO/IEC TS 33061:2021 (Process assessment model for software life cycle processes) is the general guideline for ERRM, which also uses techniques proposed by the data management maturity model (DMMM), built by Capability Maturity Model Integration Institute (CMMII). The ERMM levels and characteristics is shown in table \ref{tab:ERMM_levels_summary}.
  
  The main goal of ERMM is to leverage the quality of the emissions-related processes implemented by the cities so that the emissions information can effectively and efficiently be used in the policy-making activities towards emission reduction. The general view of ERRM is shown in Figure \ref{fig:figure_ermm_general_view}. 
  
  }
  
  \subsection{Capability Maturity Model (CMM)}

  \ifthenelse{\equal{\printtext}{true}}
  {
  
  CMM stands for capability maturity model and presents sets of recommended practices that aim to enhance software development and maintenance capabilities, as defined by \cite{CMM1993}. Thus, the CMM is built over the accumulated knowledge provided by software-process assessments and feedback from both industry and government. 
  
  As pointed out by \cite{MaturityModelsArchitecture} and \cite{ClassificationOfMaturityModels}, maturity can be considered a measure of a process related to its state or condition: defined, managed, measured, and controlled. CMM is more a set of "best practices" than a straight list of steps to be implemented. Techniques such as surveys, third-party verification, and certification\cite{CMMI2010} can be used to evaluate the level of adoption of each best practice. According to \cite{CMM2006}, the CMM is composed of five levels of maturity: initial, repeatable, defined, managed and optimized; despite the fact that the number of levels or the composition of each one it is not a rule imposed by CMM. Actually, the number of levels and what they represent can vary depending on the model to be implemented \cite{MaturityModelsArchitecture}. Figure \ref{fig:figure_cmm_general_schema} shows a general schema of CMM as defined by \cite{MaturityModelsArchitecture}. 
  
  \ifthenelse{\equal{\printfigures}{true}}
  {
  \begin{figure}[ht]
  \centering
  \includegraphics[width=\columnwidth]{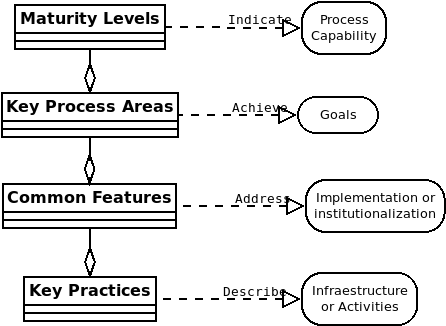}
  \caption[UML diagram of the Capability Maturity Model]{UML diagram of the Capability Maturity Model. Source: based on the Capability Maturity Model general view proposed by \cite{CMM1993}.}
  \label{fig:figure_cmm_general_schema}
  \end{figure}
  }
  
  A level in CMM is defined by analysing pre-defined capabilities that are applied to each process and its sub-processes. Some standard features and, by them, some practices are identified by analysing these processes. The evaluation also generates a list of improvements in the processes so that the next CMM level can be achieved.
  
  As an implementation example, the \cite{CMMISOTEC} defines a capability maturity model for the software development process. As expected, this model has been updated over the years, but the core components remain the same. 
  The new models derived from it and the improvements observed in the processes are a direct result of the success of this model in normalising and standardising the software development process in businesses and government organisations. 
 
  }
  
  \subsubsection{Data Management Maturity Model (DMMM)}

  \ifthenelse{\equal{\printtext}{true}}
  {

  \ifthenelse{\equal{\printfigures}{true}}
  {
  \begin{figure}[ht]
  \centering
  \includegraphics[width=\columnwidth]{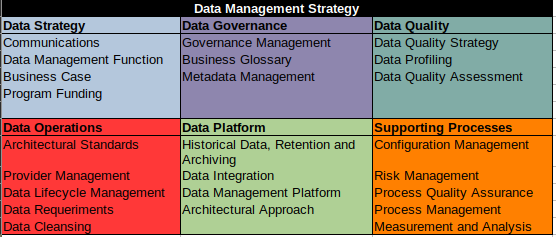}
  \caption[Data Management Maturity Model assessment example]{Data Management Maturity Model assessment example, showing numerous areas in which DMM can be applied. Source: CMMII. }
  \label{fig:figure_dmmm_assessment_view}
  \end{figure}
  }

  The Data Management Maturity Model (DMMM)\footnote{https://cmmiinstitute.com/data-management\-maturity}, built by the Capability Maturity Model Integration Institute (CMMII)\footnote{https://cmmiinstitute.com/}, is one of the derivations of CMM that focus on data management processes and their issues, in any sector and organisation, which has been more necessary than ever when organisations have to process high volumes of unstructured data daily. Applying a structured set of surveys over the business processes is one of the techniques used to implement  DMMM. In summary, the organisation has to asses the processes and data interactions within the organisation and with third parties. Figure \ref{fig:figure_dmmm_assessment_view} shows an example of a general view of the assessment and the impacted areas.
  
  }

  \subsection{ERMM in action}

  \ifthenelse{\equal{\printtext}{true}}
  {
  
  \ifthenelse{\equal{\printfigures}{true}}
  {
  \begin{figure}[ht]
  \centering
  \includegraphics[width=\columnwidth]{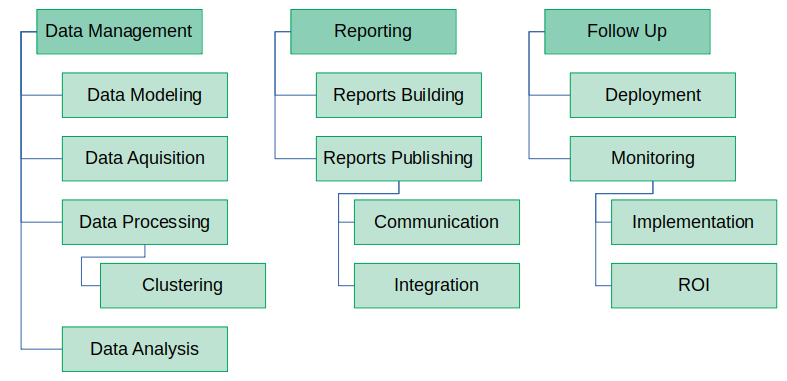}
  \caption[Emissions reporting maturity model processes view]{Emissions reporting maturity model processes view.}
  \label{fig:figure_ermm_process_view}
  \end{figure}
  }
  
  \ifthenelse{\equal{\printtables}{true}}
  {
  \begin{table*}[ht!]
  \caption[Capabilities analyzed in ERMM]{Capabilities analyzed in the context of the execution of ERMM over the cities in the CDP database. Each capability is used to evaluate and tune the evaluation of the processes present in ERMM. \\ }
  \label{tab:capabilities_to_errmlevel}
  \centering
  {\footnotesize
  \begin{tabular}{ l p{14cm} }
    \hline
    Capability & Application Example\\
    \hline
    Reliability & how reliable is the information being processed? Automated practices of data acquisition are an example of reliability level 5. \\ 
    Usability & how useful is the information to the processes. Information acquired from the available data that may compose a performance indicator is considered most useful, receiving value 5. \\
    Integration & how integrated to other sources and targets is the information. If the information is provided or validated with the help of an external source, this capability is at level 4, at least. If the channel is automated, for example, this raises to 5. \\
    Auditability & how auditable is the process and the information is treated by it. The auditability will be as good as the auditing process and resources. For example, if the information is audited by a known auditing provider with good auditing results, the level would be set to 5.\\ 
    Reproducibility & how much a process can be reproduced in other scenarios and contexts. As an example, if the process cannot be reproduced by another city because of a lack of documentation or resources, the level of capability would be set to 0. On the contrary, if conditions of reproducibility are fulfilled like human and economic resources available associated with full knowledge of the process and its pitfalls, the level would be set to 5, in this case.\\
    \hline
  \end{tabular}}
  \end{table*}
  }

  \ifthenelse{\equal{\printtables}{true}}
  {
  \begin{table*}[ht]
  \caption[Emissions Reporting Maturity Levels summary]{Emissions Reporting Maturity Levels summary. The processing contexts from which the ERM-L can be obtained are described for each level. \\ }
  \label{tab:ERMM_levels_summary}
  \centering
  {\footnotesize
  \begin{tabular}{l p{14cm}}
    \hline
    ERM Level & Contexts from which ERM level is extracted\\
    \hline
    0:Unavailable & Emissions information is not available to be used whatsoever \\
    1:Initial & Emissions information is available, but it is not part of any government plan or it cannot be validated or trusted \\   
    2:Managed & Emissions information has been used to help plan the emissions policies, but cannot be independently validated \\ 
    3:Established & Emissions information is part of the government's general plan for the city and it can be validated using in-house (local) methods \\
    4:Predictable & Emissions information is part of general and departments plans for the city and it can validate both internally and externally, by an independent auditing contractor \\
    5:Optimized & Emissions information selecting, processing and using processes are integrated into cities both short-term general and department plans and long-term policies (laws) and the actions resulting from these  can be verified independently and has their effectiveness measured. The policies derived from emissions information can also be replicated to other cities \\
    \hline
  \end{tabular}}
  \end{table*}
  }

  Based on CMM, each level of ERMM defines some goals and processes to address these goals. The processes and sub-processes are composed of practices and sub-practices that are exercised in the process implementation to be evaluated using the capabilities listed in table \ref{tab:capabilities_to_errmlevel}. The evaluation scale is from 0 (incapable) to 5 (most capable) and takes into consideration the practices being executed in the context of the evaluated process. To facilitate the evaluations, the processes are also organised in areas that indicate different contexts of application. The ERMM process schema is shown in Figure \ref{fig:figure_ermm_process_view}.
  
  A core component of ERMM is the capability evaluation matrix (CEM), which is used to associate the capabilities to the practices and sub-practices promoted by the processes being evaluated, their weights and levels of application. Figure \ref{fig:figure_ermm_evaluationmatrix_template} shows a template of it.
  
  \ifthenelse{\equal{\printfigures}{true}}
  {
  \begin{figure*}[ht]
  \centering
  \includegraphics[width=\linewidth]{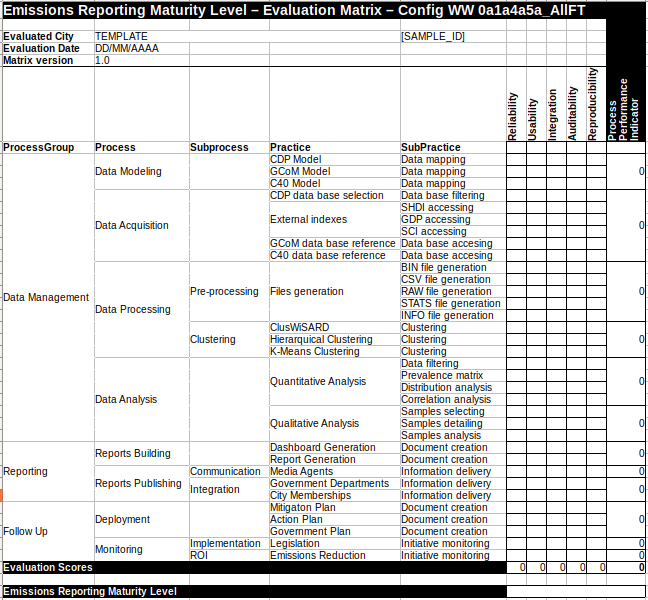}
  \caption[Emissions reporting maturity model evaluation matrix]{Emissions reporting maturity model evaluation matrix.}
  \label{fig:figure_ermm_evaluationmatrix_template}
  \end{figure*}
  }

  Another core component of ERMM is the data management context (DMC). The DMC can be defined through the analysis of a usage example based on the available data maintained by emissions awareness organisations. CDP, GCoM and C40 databases can be used to compose a virtual data model used in ERMM. One example of capability related to data sources is data modelling. In this case, the capability to model useful data to be used by emissions-related processes. The same occurs with other practices of other processes, summing the values of the performance indicators up in the execution chain. On the other side, the level of the fulfilment of the goal(s) associated with a process is also added to the performance indicator of the process.
  
  The groups of processes being evaluated are generated based on surveys oriented to find emissions-related processes or sub-processes (modules, components, pieces, or any other categorisation that indicates dependency from the parent process) among the day-to-day activities executed by companies and governments. 
  
  In an example from ERMM, the reporting group of processes is composed of construction and publication-related steps. Reports building concentrates on the generation of the document at the high administration level, using the publication of dashboards with the summary of emissions reports, and at the administrative/technical level, in which projects for future laws or mayor's decrees are built. Based on the available data, another area of interest present in the emissions reduction initiatives is the follow-up emissions policies. This group of processes deals with the ability to receive and process feedback information regarding the emissions reports applied data. At least part of the city's plans to mitigate emissions impacts should include information about emission reduction plans and emissions inventory. For example, if this information is established as effective laws, then it can help leverage the overall emissions reporting level of the city being evaluated.

  }
  
  \section{ERMM execution results}

  \ifthenelse{\equal{\printnotes}{true}}
  {
  \begin{mdframed}[leftmargin=1pt,rightmargin=1pt]
  \color{red}
  \begin{itemize}
  \item KPIs based on ERMM
  \item ERML: Emissions Reporting Maturity Level
  \item ERMLR: Emissions Reporting Maturity Level by Regions
  \end{itemize}
  \end{mdframed}
  }

  \ifthenelse{\equal{\printtext}{true}}
  {

  The Emissions Reporting Maturity Level (ERM-L) can be used to measure the overall capability of a city to select, process and deliver information about emissions in both city-wide and city-administration scopes. The ERM-L can vary from 0 to 5, as established in the emissions reporting maturity model (ERMM). The processes defined in ERMM were evaluated based on the data provided by the cities to obtain the ERM-L. The table \ref{tab:erml_execution_BR_cities} shows the ERM-L for some cities in Brazil. The PI values for the processes are also shown: data modelling, data acquisition, data processing, data analysis, build, publishing, deployment, and monitoring.
  
  }
  
  \subsection{KPI: Emissions Reporting Maturity Level}

  \ifthenelse{\equal{\printtext}{true}}
  {

  \ifthenelse{\equal{\printtables}{true}}
  {
  \begin{table*}[ht]
  \begin{threeparttable}
  \caption[ERM-L method execution for Brazil cities]{ERM-L method execution for Brazil cities.\\ }
  \label{tab:erml_execution_BR_cities}
  \centering
  {\footnotesize
  \begin{tabular}{l l c l l l l l l l l p{5cm} }
    \hline
    CDP Id	&	City Name	&	ERM-L	&	\rotatebox[origin=l]{90}{Data Modeling} &	\rotatebox[origin=l]{90}{Data Acquisition}	&	\rotatebox[origin=l]{90}{Data Processing}	&	\rotatebox[origin=l]{90}{Data Analysis}	&	\rotatebox[origin=l]{90}{Report Building}	&	\rotatebox[origin=l]{90}{Report Publishing}	&	\rotatebox[origin=l]{90}{Deployment}	&	\rotatebox[origin=l]{90}{Monitoring}	& Observations\\    
    \hline
31156	&	Curitiba	    &	1	&	1	&	110	&	0	&	1	&	0	&	0	&	1	&	1	&   \textcolor{black}{Incipient level despite high SHDI}\\
31176	&	Rio de Janeiro	&	3	&	1	&	111	&	1	&	4	&	1	&	0	&	1	&	1	&   \textcolor{black}{Possible correlation to improvements in infrastructure for international events}\\
31184	&	São Paulo	    &	2	&	1	&	101	&	1	&	1	&	1	&	0	&	1	&	1	&   \textcolor{black}{Even with higher GDP, achieved a worse result than Rio.}\\
35848	&	Belo Horizonte	&	1	&	1	&	100	&	0	&	2	&	1	&	0	&	1	&	1	&   \textcolor{black}{Incipient level, despite investments in  leveraging government administration.}\\
35865	&	Fortaleza	    &	1	&	1	&	100	&	0	&	1	&	1	&	0	&	1	&	1	&   \textcolor{black}{Incipient level, possibly correlated to north-east limitations in infrastructure}\\
35872	&	Recife	        &	0	&	0	&	100	&	0	&	2	&	0	&	0	&	1	&	1	&   \textcolor{black}{Problems with data provided  by CDP.} \\
35880	&	Porto Alegre	&	2	&	1	&	100	&	1	&	1	&	0	&	0	&	1	&	0	&   \textcolor{black}{Better than Curitiba; not SHDI related.} \\
35897	&	Campinas	    &	3	&	1	&	100	&	1	&	2	&	0	&	0	&	1	&	1	&   \textcolor{black}{Better than Sao Paulo; SHDI related.}\\
36041	&	Belém	        &	0	&	1	&	000	&	1	&	1	&	0	&	0	&	1	&	0	&   \textcolor{black}{Problems with data entering: data divergences!} \\
42120	&	Salvador	    &	1	&	1	&	110	&	0	&	1	&	1	&	0	&	1	&	1	&   \textcolor{black}{Incipient level, possibly related to north-east infrastructure limitations.} \\
42123	&	Goiânia	        &	2	&	1	&	100	&	1	&	1	&	1	&	0	&	1	&	0	&   \textcolor{black}{Agriculture recent development influence.}\\
    \hline   
  \end{tabular}}
  \begin{tablenotes}
  \item[Note:] ERM-L values can vary from 0 to 5. The range values for the performance indicators are: Data Modeling (0-1);  Data Acquisition (0-1) in each sub-item; Data Processing (0-1); Data Analysis (0-5); Report Building (0-1); Report Publishing (0-1); Deployment (0-1); Monitoring (0-1)
  \end{tablenotes}
  \end{threeparttable}
  \end{table*}
  }
  
  One of the processes evaluated to obtain the ERM-L is data acquisition. One of the practices evaluated is the quality of answers from those cities. The distribution of the quality indicator (IND) is shown in Figure \ref{fig:figure_clusters_dist_indgroup_ww0a1a4a5a_AllFT}  

  \ifthenelse{\equal{\printfigures}{true}}
  {
  \begin{figure*}[ht]
  \centering
  \includegraphics[width=\linewidth]{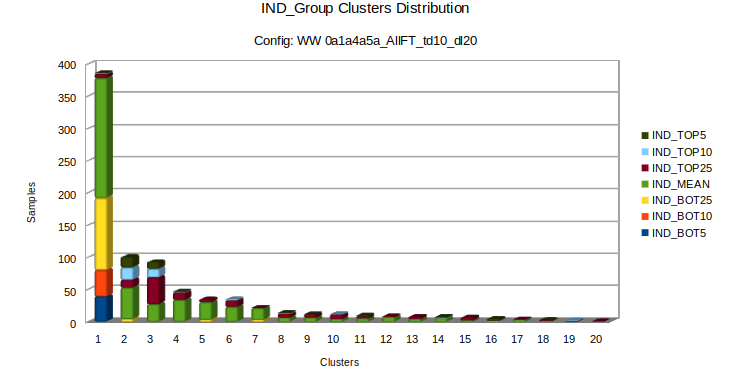}
  \caption[Quality indicator distribution using configuration WW\_0a1a4a5a\_AllFT]{Quality indicator distribution using configuration WW\_0a1a4a5a\_AllFT.}
  \label{fig:figure_clusters_dist_indgroup_ww0a1a4a5a_AllFT}
  \end{figure*}
  }
  
  Analysing the clusters distribution and quality indicator labels, clusters 5 and 7 do not have any samples in the best 10\% in terms of answering quality. It indicates the uneven balance between the answers provided by the cities and the quality of the answering process. 
 
  }

  \subsection{KPI: Emissions Reporting Maturity Level by Regions}

  \ifthenelse{\equal{\printtext}{true}}
  {

  \ifthenelse{\equal{\printnotes}{true}}
  {
  \begin{mdframed}[leftmargin=10pt,rightmargin=10pt]
  \color{red}
  \begin{itemize}
  \item Describe Emissions Reporting Maturity Level by Regions
  \end{itemize}
  \end{mdframed}
  }

  The findings obtained from the execution of the EMM-L process over the cities in the CDP database indicate differences when using the CDP region information as a filter. Further experiments executed with other region-based distributions (Country, e.g.) show similar behaviour in cluster distribution. The region-type attributes can interfere with the level of achievement of the processes and the evaluation of some capabilities. Thus, to achieve better results with ERMM, it is essential to consider region-alike attributes, even to use them in the obtained results from the method. 
  
  The distributions of quality indicator (IND) are shown in Figures \ref{fig:figure_clusters_dist_indgroup_regions1_ww0a1a4a5a_AllFT} and \ref{fig:figure_clusters_dist_indgroup_regions2_ww0a1a4a5a_AllFT}, clearly indicating the differences between the quality of the answers and the CDP regions, taking into consideration the clusters' distribution of the answers from the cities.

  \ifthenelse{\equal{\printfigures}{true}}
  {
  \begin{figure*}[ht]
  \centering
  \includegraphics[width=\linewidth]{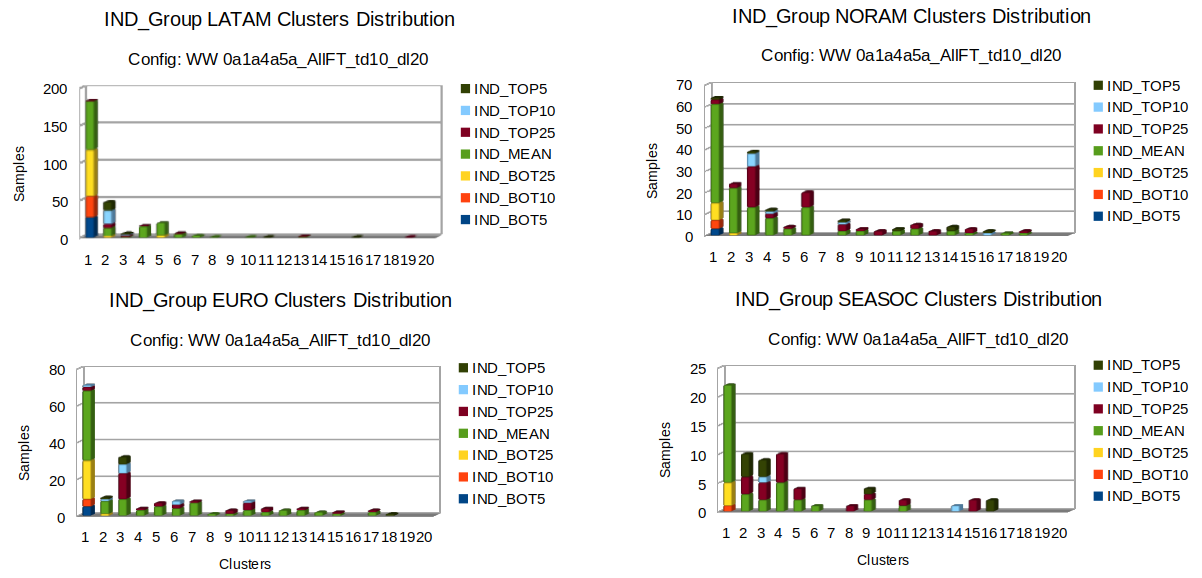}
  \caption[Quality indicator distribution using configuration WW\_0a1a4a5a\_AllFT for regions NORAM, LATAM, EURO, SEASOC]{Quality indicator distribution using configuration WW\_0a1a4a5a\_AllFT for regions NORAM, LATAM, EURO, SEASOC.}
  \label{fig:figure_clusters_dist_indgroup_regions1_ww0a1a4a5a_AllFT}
  \end{figure*}
  }
  
  \ifthenelse{\equal{\printfigures}{true}}
  {
  \begin{figure*}[ht]
  \centering
  \includegraphics[width=\linewidth]{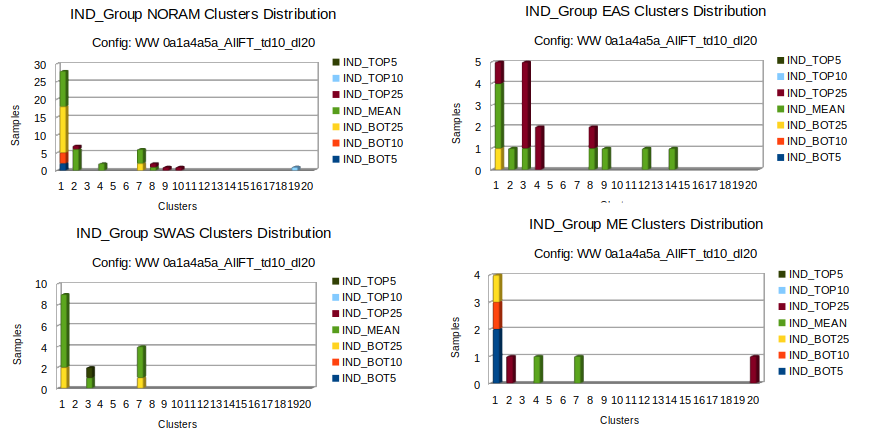}
  \caption[Quality indicator distribution using configuration WW\_0a1a4a5a\_AllFT for regions AF, SWAS, EAS, ME]{Quality indicator distribution using configuration WW\_0a1a4a5a\_AllFT for regions AF, SWAS, EAS, ME.}
  \label{fig:figure_clusters_dist_indgroup_regions2_ww0a1a4a5a_AllFT}
  \end{figure*}
  }
  
  }
  
  \section{Conclusion and future works}

  \ifthenelse{\equal{\printnotes}{true}}
  {
  \begin{mdframed}[leftmargin=1pt,rightmargin=1pt]
  \color{red}
  \begin{itemize}
  \item Emissions Reporting Maturity Model 2.0
  \item Can ERMM be applied in other areas ?
  \subitem Energy
  \subitem Transport
  \subitem Employment
  \item Can ERMM be used to implement an AI-based decision helper system ?
  \subitem 
  \end{itemize}
  \end{mdframed}
  }

  Emissions reporting empowerment is one of the highlights of this work. An effective and efficient emissions reporting system can help leverage the overall capacity of the cities to deal with emissions reduction issues and challenges. The analysed cities in this study struggle to convert emissions reporting information into actionable processes to enforce emissions reduction policies. This work points to the lack of reliable information or efficient means to correctly inform emissions facts along the decision chain as the leading cause. It also occurs when comparing the data from databases provided by cities consortia and memberships like GCoM and C40 with the disclosure data provided by the cities in the CDP database. Another issue is the absence of patterns for exchanging information about emissions-related data among the major databases, such as electronic data interchange (EDI).
  
  The performance indicators development process (PIDP) searches for PIs among the analysed data. Some correlations of the emissions reporting data and external indicators and indexes could establish the basis for PIs proposed to evaluate the cities' capacity in dealing with emissions challenges. However, the search for PIs looking into patterns extracted exclusively from the answers provided by the cities has failed, indicating a deeper problem regarding the use of available data from emissions-related studies. To confirm this hypothesis, a qualitative analysis was made based on the data produced by the clustering iterations. These analyses indicated a gap between the responses provided by the cities and the related indicators used to show emissions levels, impacts, and mitigation policies. It happened due to the low-reliability level of the information found within the sample data analysed.
  
  However, the analyses promoted in the scope of PIDP over the data could expose the inefficiencies found in the emissions reporting processes, highlighting the points in the reporting processes that can be improved. For example, consistency errors in the forms and between the information reported and external sources were constantly found in the majority of the cities. The motives for this are not established, but based on the diversity of the cities analysed that showed these difficulties, the lack of standardisation and effectiveness of the emissions reporting processes can explain that. 
  
  Thus, this work proposes the emissions reporting maturity model (ERRM)  to leverage the emissions reporting processes' efficiency and, by doing this, to achieve better results in the emissions reduction policies implementation. A city that aims to build an ERMM should apply a survey of the processes owned by the areas that deal with emissions. In this processes survey, the main goal is to identify processes impacted by or executed by emissions reduction initiatives. The survey maps procedures and operations conducted by the cities and the related processes (if they exist), the related goals of each process and the practices exercised by them. Thus, performance indicators are defined to gauge the impact of the implementation of these processes. Nevertheless, it is expected of a maturity model to have improvements over time, mainly because its effectiveness is tightly related to its application. 
  
  The findings of this work also suggest that the reporting issues associated with the emissions policies in the cities apply to other areas of interest: energy, transportation, and employment are some areas that can benefit from a reporting maturity model. The ERMM is flexible enough to embrace these other areas and their challenges. The mapped processes, goals, practices and capabilities can transcend the challenges specific to each area of interest.

  Another possible future contribution is to extend the ERMM to help design an AI-based helper system for e-government full implementation. The ERMM can map the processes that use "Internet of things" (IoT) to provide reliable information about emissions. Furthermore, the ERMM can use AI to search for patterns, best performance cases successfully applied policies and social and economic return over investment (ROI). Finally, the evolutive aspect of ERMM is an advantage to the cities to adopt and share expertise in emissions reduction policies.
  
  \bibliographystyle{elsarticle-harv.bst}

\begin{thebibliography}{5}

  \bibitem[Ratchayuda Kongboon and Shabbir H. Gheewala and Sate Sampattagul (2022)]{greenhousegasemissionsinventory}
    Ratchayuda Kongboon and Shabbir H. Gheewala and Sate Sampattagul (2022),
    \emph{Greenhouse gas emissions inventory data acquisition and analytics for low carbon cities}. Journal of Cleaner Production, 343, 130711.
    https://doi.org/10.1016/j.jclepro.2022.130711

  \bibitem[Hannah Ritchie and Max Roser and Pablo Rosado (2020)]{owidco2andgreenhousegasemissions}
    Hannah Ritchie and Max Roser and Pablo Rosado (2020),
    \emph{CO\textsubscript{2} and Greenhouse Gas Emissions},
    Our World in Data,
    https://ourworldindata.org/co2-and-greenhouse-gas-emissions

  \bibitem[R.E.H. Sims, R.N. Schock, A. Adegbululgbe, J. Fenhann, I. Konstantinaviciute, W. Moomaw (2007)]{ghgandcities}
    R.E.H. Sims, R.N. Schock, A. Adegbululgbe, J. Fenhann, I. Konstantinaviciute, W. Moomaw, et al. (2007),
    \emph{Climate Change 2007: Mitigation. Contribution of Working Group III to the Fourth Assessment Report of the Intergovernmental Panel on Climate Change}, Energy supply, Cambridge University Press, Cambridge, United Kingdom and New York, NY, USA (2007), B. Metz, O.R. Davidson, P.R. Bosch, R. Dave, L.A. Meyer (Eds.)
    
  \bibitem[Pietrapertosa, F., Salvia, M., De Gregorio Hurtado, S., D’Alonzo, V., Church, J.M.,
Geneletti, D., Reckien, D. (2019)]{urbanclimatechangemitigation}
    Pietrapertosa, F., Salvia, M., De Gregorio Hurtado, S., D’Alonzo, V., Church, J.M.,
Geneletti, D., Reckien, D. (2019),
    \emph{Urban climate change mitigation and adaptation
planning: are Italian cities ready?}. Cities 91, 93–105. https://doi.org/10.1016/j.cities.2018.11.009.

  \bibitem[A. Gouldson, S. Colenbrander, A. Sudmant, N. Godfrey, J. Millward-Hopkins, W. Fang et al (2015)]{acceleratinglowcarbondevelopment}
  A. Gouldson and S. Colenbrander and A. Sudmant and N. Godfrey and J. Millward-Hopkins and W. Fang et al (2015), 
  \emph{Accelerating Low-Carbon Development in the World's Cities. Contributing paper for Seizing the Global Opportunity: Partnerships for Better Growth and a Better Climate},
  New Climate Economy, London and Washington, DC 2015

  \bibitem[Edoardo Croci and Benedetta Lucchitta and Greet Janssens-Maenhout and Simone Martelli and Tania Molteni (2017)]{urbanco2mitigation}
  Edoardo Croci and Benedetta Lucchitta and Greet Janssens-Maenhout and Simone Martelli and Tania Molteni (2017),
  \emph{Urban CO2 mitigation strategies under the Covenant of Mayors: An assessment of 124 European cities}. Journal of Cleaner Production, v. 169, pp 161-177, https://doi.org/10.1016/j.jclepro.2017.05.165

  \bibitem[Heinonen, J., Jalas, M., Juntunen, J.K., Ala-Mantila, S., Junnila, S. (2013)]{greenhousegasimplications}
  Heinonen, J., Jalas, M., Juntunen, J.K., Ala-Mantila, S., Junnila, S. (2013)
  \emph{Situated lifestyles: I. How lifestyles change along with the level of urbanization and what the greenhouse gas implications study of Finland.} Environment Res. Lett. 8, http://dx.doi.org/10.1088/1748-9326/8/2/025003

  \bibitem[Reckien, D. and Flacke, J. and Olazabal, M. and  Heidrich, O. (2015)]{CitiesMitigationPlans}
  Reckien, D. and  Flacke, J. and Olazabal, M. and Heidrich, O. (2015),
  \emph{The influence of drivers and barriers on urban adaptation and mitigation plans: An empirical analysis of European cities.},
  PLoS One, 10(8), e0135597. https://doi.org/10.1371/journal.pone.0135597.

  \bibitem[De Gregorio Hurtado, S., Olazabal, M., Salvia, M., Pietrapertosa, F., Olazabal, E., Geneletti, D., Reckien, D. (2014)]{GovernanceStructuresInClimateActions}
  De Gregorio Hurtado, S., Olazabal, M., Salvia, M., Pietrapertosa, F., Olazabal, E., Geneletti, D., Reckien, D. (2014),
  \emph{Implications of governance structures on urban climate action: Evidence from Italy and Spain.},
  BC3 working paper series. vol. 2.

  \bibitem[Peter Smith (1990)]{performanceindicatorpublicsector}
  Peter Smith (1990),
  \emph{The Use of Performance Indicators in the Public Sector},
  Journal of the Royal Statistical Society Series A, Royal Statistical Society, vol. 153(1), pages 53-72 

  \bibitem[Fisher, N.I. (2019)]{performancemeasurment}
  Fisher, N.I. (2019),
  \emph{A comprehensive approach to problems of performance measurement},
   J. R. Stat. Soc. A, 182: 755-803. https://doi.org/10.1111/rssa.12424

  \bibitem[Smith, P.C. and Street, A. (2005)]{publicsectordata}
  Smith, P.C. and Street, A. (2005),
  \emph{Measuring the efficiency of public services: the limits of analysis},
  Journal of the Royal Statistical Society: Series A (Statistics in Society), 168: 401-417. https://doi.org/10.1111/j.1467-985X.2005.00355.x

  \bibitem[Dransfield, S. B., Fisher, N. I. and Vogel, N. J. (1999)]{stackeholdercenteredprocess}
  Dransfield, S. B., Fisher, N. I. and Vogel, N. J. (1999)
  \emph{Using statistics and statistical thinking to improve organisational performance (with discussion)}, Int. Statist. Rev., 67, 99–150
  
  \bibitem[David Parmenter (2010)]{KPI}
  David Parmenter (2010) \emph{Key performance indicators: developing, implementing, and using winning KPIs}, John Wiley \& Sons.

  \bibitem[Seyedali Mirjalili, Jin Song Dong (2019)]{OptimizationUsingAITechniques}
  Seyedali Mirjalili, Jin Song Dong (2019),
  \emph{Multi-Objective Optimization using Artificial Intelligence Techniques},
  SpringerBriefs in Applied Sciences and Technology, pages XI 58
  https://doi.org/10.1007/978-3-030-24835-2

  \bibitem[Papalexiou and Montanari(2019)]{GlobalRegionalIncreasePrecipitationExtremesUnderGlobalWarming}
  Papalexiou, S. M., Montanari, A. (2019) \emph{Global and regional increase of precipitation extremes under global warming. Water Resources Research, 55, 4901– 4914. https://doi.org/10.1029/2018WR024067} 

  \bibitem[David C. Broadstock et all (2018)]{EmissionsReportingReview}
   David C. Broadstock and Alan Collins and Lester C. Hunt and Konstantinos Vergos \emph{Voluntary disclosure, greenhouse gas emissions and business performance: Assessing the first decade of reporting, The British Accounting Review,
   vol 50, pp 48-59}

  \bibitem[Sarah Giest (2017)]{AnalyticsCarbonEmissions}
  Sarah Giest (2017), \emph{Big data analytics for mitigating carbon emissions in smart cities: opportunities and challenges, European Planning Studies, vol. 25, num. 6, pp 941-957}

  \bibitem[Kamlesh Tiwari and Mohammad Shadab Khan (2020)]{SustainabilityAccountingReporting}
  Kamlesh Tiwari and Mohammad Shadab Khan (2020), \emph{Sustainability accounting and reporting in the industry 4.0, Journal of Cleaner Production, vol. 258, num. 120783}

  \bibitem[L.P. Zhang and P. Zhou (2018)]{LowCarbonIndicator}
  L.P. Zhang and P. Zhou (2018), \emph{A non-compensatory composite indicator approach to assessing low-carbon performance, European Journal of Operational Research}

  \bibitem[Baptiste Pillain et al.(2017)]{SustainabilityIndicators}
  Baptiste Pillain and Eskinder Gemechu and Guido Sonnemann (2017), \emph{Identification of Key Sustainability Performance Indicators and related assessment methods for the carbon fiber recycling sector, Ecological Indicators, vol 72, pp 833-847}

  \bibitem[Yingli Lou et al.(2019)]{LCCIndicators}
  Yingli Lou and Wadu Mesthrige Jayantha and Liyin Shen and Zhi Liu and Tianheng Shu (2019), \emph{The application of low-carbon city (LCC) indicators - A comparison between academia and practice, Sustainable Cities and Society, vol 51, n 101677 }

  \bibitem[Paulk et al.(1993)]{CMM1993}
  Mark C. Paulk and Bill Curtis and Mary Beth Chrissis and Charles V. Weber (1993), \emph{Capability maturity model, version 1.1, IEEE Software}

  \bibitem[Monteiro and Maciel (2020)]{MaturityModelsArchitecture}
  Erasmo L. Monteiro and Rita S. Pitangueira Maciel (2020) \emph{Maturity Models Architecture: A large systematic mapping, Revista Brasileira de Sistemas de Informação (Brazilian Journal of Information Systems), vol 13, num 12, pp 110-140} 

  \bibitem[Mettler et al. (2010)]{ClassificationOfMaturityModels}
  Tobias Mettler and Peter Rohner and Robert Winter (2010) \emph{Towards a Classification of Maturity Models in Information Systems, Management of the Interconnected World, pp 333-340}

  \bibitem[Capability Maturity Model Institute (2010)]{CMMI2010}
  Capability Maturity Model Institute (2010)\emph{CMMI for Development, Version 1.3, SEI-2010-TR-033}

  \bibitem[Capability Maturity Model Institute (2006)]{CMM2006}
  Mark C. Paulk and Bill Curtis and Mary Beth Chrissis and Charles V. Weber (2006)
  \emph{Capability maturity model, version 1.2, IEEE Software, vol 10, num 4, pp 18 - 27}

  \bibitem[International Organization for Standardization (2021)]{CMMISOTEC}
  International Organization for Standardization \emph{Process assessment model for software life cycle processes,TS-33061}

  \bibitem[Liebchen and Shepperd (2016)]{DataSetAndDataQuality}
  Liebchen, Gernot and Shepperd, Martin (2016), \emph{Data Sets and Data Quality in Software Engineering: Eight Years On, Proceedings of the The 12th International Conference on Predictive Models and Data Analytics in Software Engineering, PROMISE 2016, Association for Computing Machinery, vol 122, pp 7:1-7:4 }

  \bibitem[Booch et al.(2005)]{UML}
  Grady Booch and James Rumbaugh and Ivar Jacobson (2005), \emph{Unified Modeling Language User Guide, The, 2nd Edition, Addison-Wesley Professional}

  \bibitem[Shepperd et al.(2013)]{DataQualityNASA}
  Martin Shepperd and Qinbao Song and Zhongbin Sun and Carolyn Mair (2013), \emph{Data Quality: Some Comments on the NASA Software Defect Datasets, IEEE TRANSACTIONS ON SOFTWARE ENGINEERING, vol. 39, n. 9, pp. 1208-1215}

  \bibitem[Zhao and Sahni(2017)]{StringCorrectionUsingDL}
  Zhao and Sahni(2017),
 \emph{String correction using the Damerau-Levenshtein distance (2017), 7th IEEE International Conference on Computational Advances in Bio and Medical Sciences (ICCABS 2017), Orlando, FL, USA, pp. 20-47}

  \bibitem[Ben J. Clarke and Friederike and Richard G. Jones(2021)]{CLARKE2021100285}
  Ben J. Clarke and Friederike and Richard G. Jones(2021),
\emph{Inventories of extreme weather events and impacts: Implications for loss and damage from and adaptation to climate extremes (2021), Journal of Climate Risk Management, vol 32, pp. 100285}

  \bibitem[Yousef Sangsefidi and Austin Barnes and Mark Merrifield and Hassan Davani(2023)]{SANGSEFIDI2023104914}
  Yousef Sangsefidi and Austin Barnes and Mark Merrifield and Hassan Davani(2023),
  \emph{Data-driven analysis and integrated modeling of climate change impacts on coastal groundwater and sanitary sewer infrastructure (2023), Journal of Sustainable Cities and Society, vol 99, pp. 104914}

  \bibitem[Giulia Datola(2023)]{DATOLA2023104821}
  Giulia Datola(2023),
  \emph{Implementing urban resilience in urban planning: A comprehensive framework for urban resilience evaluation (2023), Journal of Sustainable Cities and Society, vol 98, pp. 104821}

  \bibitem[Diogo Cunha Ferreira, José Rui Figueira, Salvatore Greco, Rui Cunha Marques(2022)]{DEAWithInperfectData}
  Diogo Cunha Ferreira, José Rui Figueira, Salvatore Greco, Rui Cunha Marques(2023),
  \emph{Data Envelopment Analysis models with imperfect knowledge of input and output values: An application to Portuguese public hospitals (2023), Journal of Expert Systems with Applications, vol 231, pp. 120543}

  \bibitem[Muhammad Bilal, Lukumon O. Oyedele(2020)]{BigDataWithMLandKPI}
  Muhammad Bilal, Lukumon O. Oyedele(2020),
  \emph{Big Data with deep learning for benchmarking profitability performance in project tendering (2020), Journal of Expert Systems with Applications, vol 147, pp. 113194}

  \bibitem[Riccardo Patriarca, Francesco Simone, Giulio Di Gravio(2022)]{WheaterForecastingWithMLandHierarchicalClustering}
  Riccardo Patriarca, Francesco Simone, Giulio Di Gravio(2022),
  \emph{Supporting weather forecasting performance management at aerodromes through anomaly detection and hierarchical clustering(2022), Journal of Expert Systems with Applications, vol 213, part C, pp. 119210}

  \bibitem[Bojan Srdjevic, Zorica Srdjevic(2023)]{AnalyticHierachyProcessWithMatrices}
  Bojan Srdjevic, Zorica Srdjevic(2023),
  \emph{Prioritisation in the analytic hierarchy process for real and generated comparison matrices(2023), Journal of Expert Systems with Applications, vol 225, pp. 120015}

  \end{thebibliography}

\end{document}